\documentclass[acmsmall,screen, nonacm]{acmart}
\usepackage{hyperref}
\usepackage{lineno}
\usepackage{enumitem}
\usepackage{adjustbox}
\usepackage{multirow}
\usepackage{subcaption}
\usepackage{tablefootnote}
\usepackage{float}
\usepackage{comment}
\usepackage{xcolor}
\usepackage[normalem]{ulem}
\usepackage[ruled,lined,boxed,linesnumbered]{algorithm2e}
\usepackage{url}
\usepackage{graphics}
\usepackage[defaultlines=3,all]{nowidow}
\usepackage{tikz}
\usepackage{subcaption}
\usepackage{tabularx}
\usepackage{booktabs}
\usepackage{siunitx}
\usepackage{graphicx}
\usepackage{listings}
\usepackage{float}
\usepackage{comment}
\usepackage{amsmath,amsfonts}
\usepackage{array}
\usepackage{booktabs}      
\usepackage{siunitx}       
\newcommand{\cmark}{\checkmark}
\newcommand{\xmark}{\text{\sffamily X}}
\usepackage{textcomp}
\usepackage{stfloats}
\usepackage{verbatim}
\usepackage{graphics}
\usepackage{tikz}
\usepackage{listings}
\usepackage{graphicx}
\usepackage{subcaption}
\usepackage{makecell}
\usepackage{algpseudocode}
\usepackage[all]{nowidow}

\usepackage{mathtools}
\usepackage{algpseudocode}

\usepackage{tikz}
\usepackage{nowidow}
\usepackage{listings}
\usepackage{float}
\usepackage{comment}
\usepackage[ruled,lined,boxed,linesnumbered]{algorithm2e}

\SetKwInput{KwOutput}{Output}
\SetKwInput{KwInput}{Input}
\newcommand{\nosemic}{\renewcommand{\@endalgocfline}{\relax}}
\newcommand{\dosemic}{\renewcommand{\@endalgocfline}{\algocf@endline}}
\let\oldnl\nl
\newcommand{\nonl}{\renewcommand{\nl}{\let\nl\oldnl}}
\definecolor{mypurple}{RGB}{128,0,128}
\definecolor{mybrown}{RGB}{165,42,42}

\newcounter{reviewercount}
\newcounter{commentcount}
\newcommand{\reviewer}{\bigskip\noindent {\bf COMMENTS OF REVIEWER
    \#\refstepcounter{reviewercount}\thereviewercount.}
    \setcounter{commentcount}{0}\par\rm\em
}

\newcommand{\aeditor}%
  {\bigskip\noindent {\bf COMMENTS OF THE ASSOCIATE EDITOR}%
  \setcounter{commentcount}{0}\par 
}
\newcommand{\rcomment}{\bigskip\par\noindent{\bf Comment
   \refstepcounter{commentcount}
   \thereviewercount.\thecommentcount: }\rm\em
}
\newcommand{\response}{\smallskip\par\noindent{\bf Response: }\em\rm }

\acmJournal{TODAES}

\begin{document}

\setcounter{page}{1}
\pagestyle{standardpagestyle}
\title{ChipletPart: Cost-Aware Partitioning for 2.5D Systems}
\author{Alexander Graening}
\email{agraening@ucla.edu }
\orcid{0009-0008-2150-3738}
\author{Puneet Gupta}
\email{puneetg@ucla.edu}
\orcid{0000-0002-6188-1134}
\affiliation{
    \institution{University of California, Los Angeles}
    \city{Los Angeles}
    \state{California}
    \country{USA}
}
\author{Andrew~B.~Kahng}
\email{abk@ucsd.edu}
\orcid{0000-0002-4490-5018}
\author{Bodhisatta Pramanik}
\email{bopramanik@ucsd.edu}
\orcid{0009-0004-6014-1048}
\author{Zhiang Wang}
\email{zhw033@ucsd.edu}
\orcid{0000-0002-6669-9702}
\affiliation{
    \institution{University of California, San Diego}
    \city{La Jolla}
    \state{California}
    \country{USA}
}

\renewcommand{\shortauthors}{Graening et al.}

\begin{abstract}
    Industry adoption of chiplets has been \textcolor{black}{growing} as \textcolor{black}{chiplets} \textcolor{black}{are} a cost-effective option for making large, high-performance systems.
    Consequently, partitioning large systems into chiplets is increasingly important. In this work, we introduce \emph{ChipletPart} --- a cost-driven 2.5D system partitioner that addresses the unique constraints of chiplet systems, including complex objective functions, limited reach of inter-chiplet I/O transceivers, and the assignment of heterogeneous manufacturing technologies to different chiplets. \emph{ChipletPart} integrates a sophisticated chiplet cost model with \textcolor{black}{a} genetic algorithm \textcolor{black}{(GA)}-based technology assignment and partitioning methodology, along with a simulated annealing \textcolor{black}{(SA)}-based chiplet floorplanner. Our results show that \textcolor{black}{{\em ChipletPart}}:
        (i) \textcolor{black}{reduces} chiplet cost by up to 58\% (20\% geometric mean) compared to state-of-the-art min-cut partitioners, which often yield floorplan-infeasible solutions;
        (ii) \textcolor{black}{generates} partitions with up to 47\% (6\% geometric mean) lower cost \textcolor{black}{compared to} the prior work \emph{Floorplet};
        (iii) \textcolor{black}{reduces} chiplet cost up to 48\% (30\% geometric mean) compared to {\em Chipletizer}, while consistently producing I/O-feasible chiplet solutions across all testcases; and
        (iv) for the testcases we study, heterogeneous integration reduces cost by up to \textcolor{black}{43\%} (15\% geometric mean) compared to homogeneous implementations. Additionally, we explore Bayesian optimization (BO) for finding low cost and floorplan-feasible chiplet solutions with technology assignments. On some testcases, our BO framework achieves better \textcolor{black}{system} cost (up to 5.3\% improvement) with higher runtime overhead (up to 4$\times$) compared to our GA\textcolor{black}{-based} framework. We also present case studies that show how changes in packaging and inter-chiplet signaling technologies can affect partitioning solutions. Finally, \emph{ChipletPart}, the underlying chiplet cost model, and \textcolor{black}{our} chiplet testcase generator \textcolor{black}{are} available as open-source tools for the community.
\end{abstract}



\maketitle

\section{Introduction}
\label{sec:intro}

The integration of multiple chips on an interposer has become a 
favorable approach to reduce the cost of 
building large systems \textcolor{black}{\cite{IntelPV, Sangiovanni-VincentelliLZZ23}}. 
In a 2.5D system, a design is decomposed 
into multiple smaller chiplets, which are packaged together on 
the same substrate. This splitting can have significant 
benefits for yield and can enable designs using heterogeneous 
process technology nodes, going beyond what is possible in 
a monolithic design. On the other hand, inter-die communication 
exhibits higher area and power overhead compared to intra-die 
communication. \textcolor{black}{Consequently}, the potential cost 
benefits of disaggregation are not always realized 
\cite{GraeningPG23}. Therefore, it is important to intelligently 
partition a design into constituent chiplets 
{\em and} assign a manufacturing technology to each \textcolor{black}{chiplet} so as to optimize manufacturing costs.

\begin{figure}
    \centering
    \includegraphics[width=0.8\columnwidth]{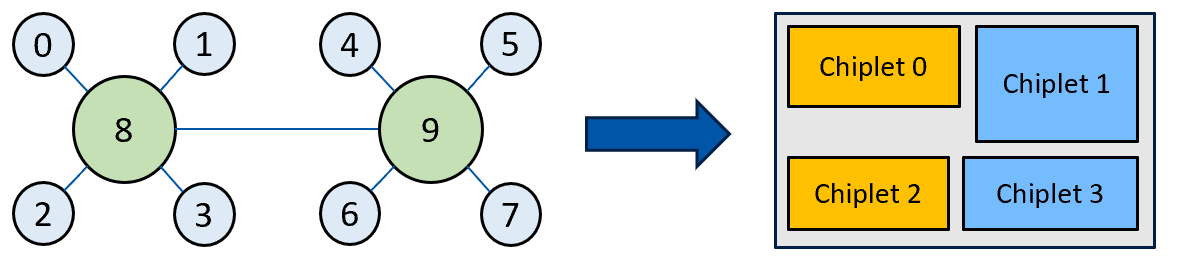}
    \Description{An image showing colored circles on the left representing blocks being partitioned into colored rectangles representing chiplets on the right.}
    \caption{Generic example of block-level chiplet partitioning. 
    Left: an IP block-level netlist (different types of IP blocks 
    shown in different colors). Right: an integrated 2.5D chiplet 
    system reflecting  \textcolor{black}{the block-level} netlist partitioning with technology node
    assignment shown  \textcolor{black}{in} different colors.}
    \label{fig:chiplet_diagram}
\end{figure}

The natural partitioning granularity for 2.5D integration is at 
the block-level. Splitting individual IP blocks 
into multiple chiplets is problematic for design and test 
methodology reasons as well as for IP reuse. Additionally, 
splitting an IP block will often have a significant impact on 
performance and \textcolor{black}{I/O} count. Due to these considerations, 
block-level chiplet partitioning (Figure 
\ref{fig:chiplet_diagram}) is the focus of this paper. 
The cost-aware partitioning 
problem for chiplet systems is fundamentally different from the 
well-studied netlist min-cut partitioning~\textcolor{black}{\cite{KarypisAKS99, bustany2022specpart, BustanyGKKPW23}}. \textcolor{black}{For the former}, the problem 
size is smaller (usually a few hundred blocks and a few 
tens of chiplets at most) and the underlying \textcolor{black}{objective function} can be 
complex. Furthermore, inter-chiplet \textcolor{black}{I/O} reach limitations 
can render some partitions infeasible, necessitating 
floorplan-awareness in the partitioner. This is further 
complicated by the need to select a manufacturing technology 
for each \textcolor{black}{chiplet}.

\textcolor{black}{We propose {\em ChipletPart}, the first open-source, 
unified framework for chiplet partitioning that combines {\em cost-driven} multi-way 
partitioning, heterogeneous technology assignment and \textcolor{black}{I/O} {\em reach-aware} 
floorplanning in a single, automated flow. 
{\em ChipletPart} presents a novel adaptation of classical optimizers such as genetic 
algorithms  \textcolor{black}{(GA)} and
simulated annealing \textcolor{black}{(SA)} to generate floorplan-feasible, cost-optimized \textcolor{black}{2.5D chiplet system partitions with technology assignments.} 
{\em ChipletPart} is \textcolor{black}{intended for design space exploration before physical implementation}.\footnote{\textcolor{black}{{\em ChipletPart} considers
a given architecture and generates partitioned chiplets before physical implementation.} \textcolor{black}{We do not perform architecture exploration in this work. We envision {\em ChipletPart} as a tool to assist human
designers by generating strong initial solutions, which can then be
refined using domain expertise.}} Our key contributions include:}

\begin{itemize}[noitemsep,topsep=0pt,leftmargin=*]
    \item \textcolor{black}{\textbf{Integrated chiplet partitioning with \textcolor{black}{floorplan-feasibility}:} We present 
    the first unified framework that optimizes system partitioning 
    while guaranteeing feasibility under \textcolor{black}{I/O} reach constraints. 
    In contrast to standard 
    min-cut partitioners (e.g., {\em hMETIS} \cite{KarypisAKS99}, {\em TritonPart} \cite{BustanyGKKPW23}) that ignore \textcolor{black}{floorplan-feasibility}, {\em ChipletPart} employs a reach-aware, \textcolor{black}{simulated} annealing-based placement engine to guarantee \textcolor{black}{floorplan-feasible} solutions (Section~\ref{sec:floorplan}).
    }
    \item \textcolor{black}{\textbf{Heterogeneous technology assignment:} We handle technology node assignment 
    during partitioning using a genetic algorithm. This enables exploration of cost trade-offs in heterogeneous 2.5D systems --- a capability not supported by existing tools such as \textcolor{black}{{\em Floorplet}~\cite{ChenLZZL23}} or \textcolor{black}{{\em Chipletizer}~\cite{Li24}}. Our  
    heterogeneous technology-aware, cost-driven, multi-way 
    partitioner, {\em ChipletPart}, finds cost reductions of up 
    to \textcolor{black}{$43\%$} (\textcolor{black}{6\% geometric mean}) in multi-technology scenarios. 
    To the best of our knowledge, we are the first work 
    that does technology assignment during partitioning (Section~\ref{sec:approach}).} \textcolor{black}{In addition, we 
    explore the use of Bayesian optimization (BO) to search for 
    floorplan-feasible, cost-\textit{optimal} chiplet solutions with 
    technology assignments. While BO improves system cost by up to 5.3\% 
    over our GA-based framework on a few testcases, it incurs a runtime 
    overhead of up to $4\times$. Thus, GA and BO provide \textcolor{black}{a quality-runtime 
    trade-off} (Section~\ref{subsec:bayesian_opt}).}

    \item \textbf{Improvements over SOTA:} {\em ChipletPart} achieves up to 46\% improvement in cost over state-of-the-art min-cut based partitioners ({\em hMETIS} \cite{KarypisAKS99}, {\em TritonPart} \cite{BustanyGKKPW23}) that do not guarantee \textcolor{black}{floorplan-feasibility}. Compared to the {\em parChiplet} partitioner from {\em Floorplet}~\cite{ChenLZZL23}, {\em ChipletPart} generates chiplet solutions that are up to $47\%$ (6\% geometric mean) better in cost. Compared to manual partitions, {\em ChipletPart} generates $34\%$ (13\% geometric mean) better solutions (Section~\ref{sec:results}). We also compare against the cost-driven {\em Chipletizer}~\cite{Li24} framework and observe that {\em ChipletPart} consistently produces I/O-feasible partitions with up to 48\% lower cost across our design suite (Section~\ref{sec:results}). Additionally, we explore how variations in technology parameters affect chiplet partitioning (Section~\ref{sec:case_studies}).

    \item \textbf{Standardized open-source ecosystem:} We \textcolor{black}{translated} 
    the Python-based cost 
    model \cite{Graening_CATCH, costmodelUCLA} to an 
    equivalent C++ implementation 
    (Section~\ref{sec:cost}). This yields a $5\times$ speedup 
    in cost model execution runtime, and a speedup of more than 
    $100\times$ in our \textcolor{black}{multithreaded} implementation. {\em ChipletPart}, \textcolor{black}{along} with its core cost model, 
    is permissively open-sourced~\cite{anonymousrepo}, 
    enabling others to readily adapt it for benchmarking and 
    further development. Also, we make our testcases 
    publicly available, enabling the community to access a new, 
    standardized set of benchmarks.

\end{itemize}

\noindent

In the following, 
Section \ref{sec:related_works} reviews related works
on chiplet cost modeling, partitioning and floorplanning. 
Sections \ref{sec:approach} and
\ref{sec:partition} respectively provide an overview and details of 
our partitioning approach.
Section \ref{sec:cost_driver} describes the driving
considerations in chiplet partitioning. Sections \ref{sec:results} 
and \ref{sec:case_studies} show experimental results 
and additional case studies, and Section \ref{sec:conclusion}
concludes the paper. 

\section{Related Work}
\label{sec:related_works}

We now discuss fundamentals and previous works on 
chiplet cost modeling,
then review existing works on chiplet 
partitioning and floorplanning.
Tables \ref{tab:general_terms}  and  \ref{tab:ga_specific_terms} 
summarize key terms and notations.

\subsection{Chiplet Cost Modeling}
\label{sec:related_cost}

Cost reduction is one of the main drivers for developing
2.5D systems. The total cost of developing a VLSI system can be 
divided into two parts: non-recurring engineering 
(NRE) cost and recurring engineering (RE) cost \cite{Feng22}.
NRE cost refers to the one-time cost of designing a VLSI system, 
including IP qualification, architecture \textcolor{black}{simulation}, 
verification, physical design, software \textcolor{black}{license fees}, etc.
RE cost refers to the fabrication costs in mass production, 
such as \textcolor{black}{wafers, assembly, and test.}
Researchers have proposed several chiplet cost models to estimate 
various components of the total 2.5D system cost.
\cite{Feng22} introduces a quantitative cost model for comparing 
RE and NRE costs between monolithic SoC and 
multi-chip integration, \textcolor{black}{but accounts for fewer RE cost considerations than our chosen model.}
\cite{ZhuangYCH22} and \cite{ChenLZZL23} propose cost models that 
take into account reliability issues such as
bump stress and warpage.
\cite{GraeningPG23, Graening_CATCH} \textcolor{black}{present} a case study of a 
large system built using chiplets and \textcolor{black}{analyze} the sensitivity of 
system cost to factors such as defect density, assembly cost, \textcolor{black}{I/O} 
size, etc. The authors have made their cost 
model publicly available at~\cite{costmodelUCLA}. 
In this work, we use the cost model proposed by 
\cite{costmodelUCLA} (see Section~\ref{sec:cost_driver}) \textcolor{black}{due to the wide range of 
configurable parameters available for system technology co-optimization (STCO).\footnote{To the best of 
our knowledge, the cost model in~\cite{costmodelUCLA} is the 
most detailed open-source cost model currently available
for chiplets \textcolor{black}{at the time of this publication}.} \textcolor{black}{The cost model in} \cite{Graening_CATCH} was used in collaboration with IMEC to perform an STCO study in \cite{Graening24}. While the specific 
parameter settings in the cost model do influence partitioning outcomes, all parameters are externally 
configurable via user-defined configuration files. To enable efficient integration with our
partitioning framework, we ported the cost model to C++, resulting in significantly improved 
performance.}

\begin{table}[!t]
  \centering
  \caption{General terminology and notation.}
  \begin{tabular}{|>
  {\raggedright\arraybackslash}p{0.22\columnwidth}|>
  {\raggedright\arraybackslash}p{0.68\columnwidth}|}
    \hline
    \textbf{Notation} & \textbf{Description}  \\ \hline \hline
    $\mathcal{C}$ & Set of chiplets \\ \hline 
    $\mathcal{C}_t$ & Set of chiplets implemented in technology node $t$ \\ \hline 
    $N$ & Chiplet-level netlist \\ \hline
    $\mathcal{S}$ & Block-level netlist \\ \hline
    $V$ & Chip manufacturing volume \\ \hline
    $\mathcal{T}$ & Set of technology nodes \\ \hline
    $\omega$ & Mapping from chiplets to technology nodes \\ \hline
    $\phi$ & Partitioning \textcolor{black}{objective} function \\ \hline
    $m$ & Number of different technology nodes \\ \hline
  \end{tabular}
  \label{tab:general_terms}
\end{table}

\begin{table}
  \centering
  \caption{GA terminology and notation. \textcolor{black}{Default hyperparameters
  were empirically determined (Section~\ref{subsubsec:hyperparam_explore}).}}
    \begin{tabular}{|>
    {\raggedright\arraybackslash}p{0.17\columnwidth}|>
    {\raggedright\arraybackslash}p{0.83\columnwidth}|}
    \hline
    \textbf{Notation} & \textbf{Description}  \\ \hline \hline
    $tot_{pop}$ & \textcolor{black}{ \textcolor{black}{Number of} members in the population} 
    \textcolor{black}{(default 50)} \\ \hline
    \textcolor{black}{$k_{pop}$} & \textcolor{black}{Number of 
    pairs of parents selected from a population for tournament 
    selection (default 45)} \\ \hline
    \textcolor{black}{$\zeta$} & \textcolor{black}{Tournament size  (default 3)} \\ \hline
    \textcolor{black}{$\sigma$} & \textcolor{black}{ \textcolor{black}{Number of} members picked for elitism (default 5)} \\ \hline
    \multirow{1}{*}{\makecell{\textcolor{black}{$\Delta_{threshold}$}}} 
    & \textcolor{black}{Threshold value for improvement  between} 
    \textcolor{black}{successive generations (default 0.01)} \\ \hline 
    \textcolor{black}{$\Psi$} & \textcolor{black}{Maximum number of generations (default 50)} \\ \hline 
     \multirow{2}{*}{\makecell{\textcolor{black}{$\epsilon$}}} & \textcolor{black}{Maximum number of successive generations with
    cost improvement less than $\Delta_{threshold}$ (default 10)} \\ \hline
    \textcolor{black}{$p_c$} & \textcolor{black}{Crossover probability (\textcolor{black}{default 0.60})} \\ \hline
    \textcolor{black}{$p_m$} & \textcolor{black}{Mutation probability (\textcolor{black}{default 0.07})} \\ \hline
    \textcolor{black}{$K_{max}$} & \textcolor{black}{Maximum number of attainable chiplets (default 8)} \\ \hline
  \end{tabular}
  \label{tab:ga_specific_terms}
\end{table}

\subsection{Chiplet Partitioning and Floorplanning}
\label{sec:related_partition}

Several previous works address chiplet partitioning. 
\cite{Kabir20} incorporates the min-cut partitioner {\em hMETIS} 
\cite{KarypisK96} within a chiplet implementation flow.
\cite{ChenLZZL23} proposes {\em parChiplet}, which partitions the 
SoC system into chiplets based on functional 
and area characteristics of IP blocks. 
However, \textcolor{black}{\cite{Kabir20, ChenLZZL23}} do not optimize the 
actual \textcolor{black}{chiplet cost}. 
{\em Chipletizer} \cite{Li24} proposes a unified design 
characterization graph
to represent both SoC designs and chiplets, and uses 
simulated annealing (SA) to optimize the overall cost of 
chiplet-based systems.
However, {\em Chipletizer} does not consider floorplan 
feasibility or \textcolor{black}{support} heterogeneous integration. 
\textcolor{black}{The recent work of \cite{VicharraCH25} uses 
reinforcement learning (RL) and
simulated annealing to perform PPA-oriented chiplet 
partitioning. \textcolor{black}{However,} their
method does not guarantee \textcolor{black}{floorplan-feasibility}.}
The main differences between our {\em ChipletPart} 
and previous works 
are summarized in Table \ref{tab:previous_works}.
\textcolor{black}{In our experimental evaluations, we compare 
{\em ChipletPart} with ~\cite{ChenLZZL23} and~\cite{KarypisK96}. 
The authors of \textcolor{black}{\cite{Li24, Kabir20, VicharraCH25}} \textcolor{black}{have not released} their source code or
binaries.\footnote{\textcolor{black}{We attempted to contact the authors of~\cite{VicharraCH25}, but \textcolor{black}{have not received a response.}}}}

\begin{table}
    \caption{\small
    \textcolor{black}{Comparison of  \textcolor{black}{state-of-the-art} chiplet partitioning methods.}}
    \label{tab:previous_works}
    \centering
    \begin{tabular}{|c|c|c|c|}
    \hline
     \makecell{\textbf{Methods}} & \makecell{\textbf{Cost-Driven}} & \makecell{\textbf{Floorplan} \\ \textbf{Feasibility}} & \makecell{\textbf{Heterogeneous} \\ \textbf{Integration}} \\    
     \hline
     \cite{ChenLZZL23}, \cite{Kabir20}, \cite{KarypisK96}  & & & \\ \hline
    \cite{Li24} & \checkmark & & \\ \hline
    \cite{VicharraCH25} & \checkmark &  & \checkmark \\ \hline
    {\em ChipletPart} & \checkmark &  \checkmark & \checkmark  \\ \hline
    \end{tabular} 
\end{table}

Existing chiplet floorplanning approaches 
fall into three categories: simulated 
annealing-based, branch-and-bound (B\&B)-based and 
mathematical programming (MP)-based. 
Works such as  \textcolor{black}{\cite{HoC13, LeeC23}} use classical 
floorplan representations
such as  B$^*$ tree and apply SA to optimize \textcolor{black}{an objective} function. 
 \textcolor{black}{\cite{ChiouJCLP23, LiuCW14}} enumerate possible 
floorplanning solutions and apply B\&B to find a near-optimal solution.
 \textcolor{black}{\cite{ChenLZZL23, ZhuangYCH22}} develop MP-based formulations 
of chiplet floorplanning, enabling MP solvers to find optimal solutions. 
We note that existing works fail to handle the ``reach'' 
constraints imposed by \textcolor{black}{I/O} cells, and generally assume that the 
size and shape of chiplets are predetermined by designers.  
 \textcolor{black}{In this work, we} propose a reach-aware chiplet floorplanner 
\textcolor{black}{that} simultaneously determines the location and 
shape for each chiplet.

\section{Our Approach}
\label{sec:approach}

We now introduce the chiplet partitioning problem, then 
give an overview of our chiplet partitioning framework.

\subsection{Problem Formulation}
\label{subsec:part_problem}

\textcolor{black}{The chiplet partitioning problem that we address} is fundamentally different from classical 
min-cut partitioning.
The key differences are summarized as follows.
\begin{itemize}[noitemsep,topsep=0pt,leftmargin=*]
    \item \textbf{Problem size}: Min-cut partitioning is 
    typically performed on gate-level netlists
    comprising millions of gates. By contrast, \textcolor{black}{we} focus on a block-level netlist that 
    typically comprises several hundreds of IP blocks.
    \item \textbf{Objective:} Min-cut partitioning focuses on 
    minimizing {\em cutsize}, the number of nets (hyperedges) 
    crossing partition boundaries. 
    By contrast, \textcolor{black}{we seek} to reduce 
    2.5D system development cost, which brings a more 
    complex set of considerations (Section ~\ref{sec:cost_driver}).
    \item \textbf{Constraints:} Min-cut partitioning minimizes
    cutsize subject to a given balance 
    constraint \cite{BustanyGKKPW23}. By contrast,
    chiplet partitioning is not necessarily balanced,
    but it must produce {\em I/O-feasible} solutions (Section \ref{sec:floorplan}).
    \item \textbf{Heterogeneity:} Min-cut partitioning is 
    insensitive to partition labels: swapping the contents
    of two partitions will not affect cutsize. 
    By contrast, chiplet partitioning is sensitive to partition
    labels (i.e., technology node assignments): swapping the
    technologies of two chiplets can significantly change total 
    cost of a 2.5D system. 
\end{itemize}
These fundamental differences between min-cut partitioning and 
chiplet partitioning motivate our studies. Formally, the inputs 
to our multi-technology chiplet partitioning framework
are a block-level netlist $ \mathcal{S} $ and a set of technology 
nodes $ \mathcal{T} $. 
The outputs are a set of chiplets $ \mathcal{C} $
and their corresponding technology 
assignment $\omega: \mathcal{C} \to \mathcal{T}$.
The objective is to minimize the total cost of 
a 2.5D system:
\begin{equation}
   \label{eq:part_cost}
    \phi_{overall}(\mathcal{C}) = \frac{\phi_{assembly} + \sum \frac{\phi_{die}}{Y_{die}}}{Y_{assembly}} + \frac{g(\omega)}{V}
\end{equation}

\noindent
where $\phi_{die}$ is the silicon cost of each chiplet, 
$\phi_{assembly}$ is the assembly cost, $Y_{die}$ is the die 
yield, 
$Y_{assembly}$ is the assembly yield, $g(\omega)$ is the 
non-recurring engineering  \textcolor{black}{(NRE)} cost, and $V$ is the chip manufacturing 
volume. The constraint is the I/O-feasibility of the chiplet set  
$\mathcal{C} $.

\begin{figure}
    \centering
    \includegraphics[width=0.8\columnwidth]{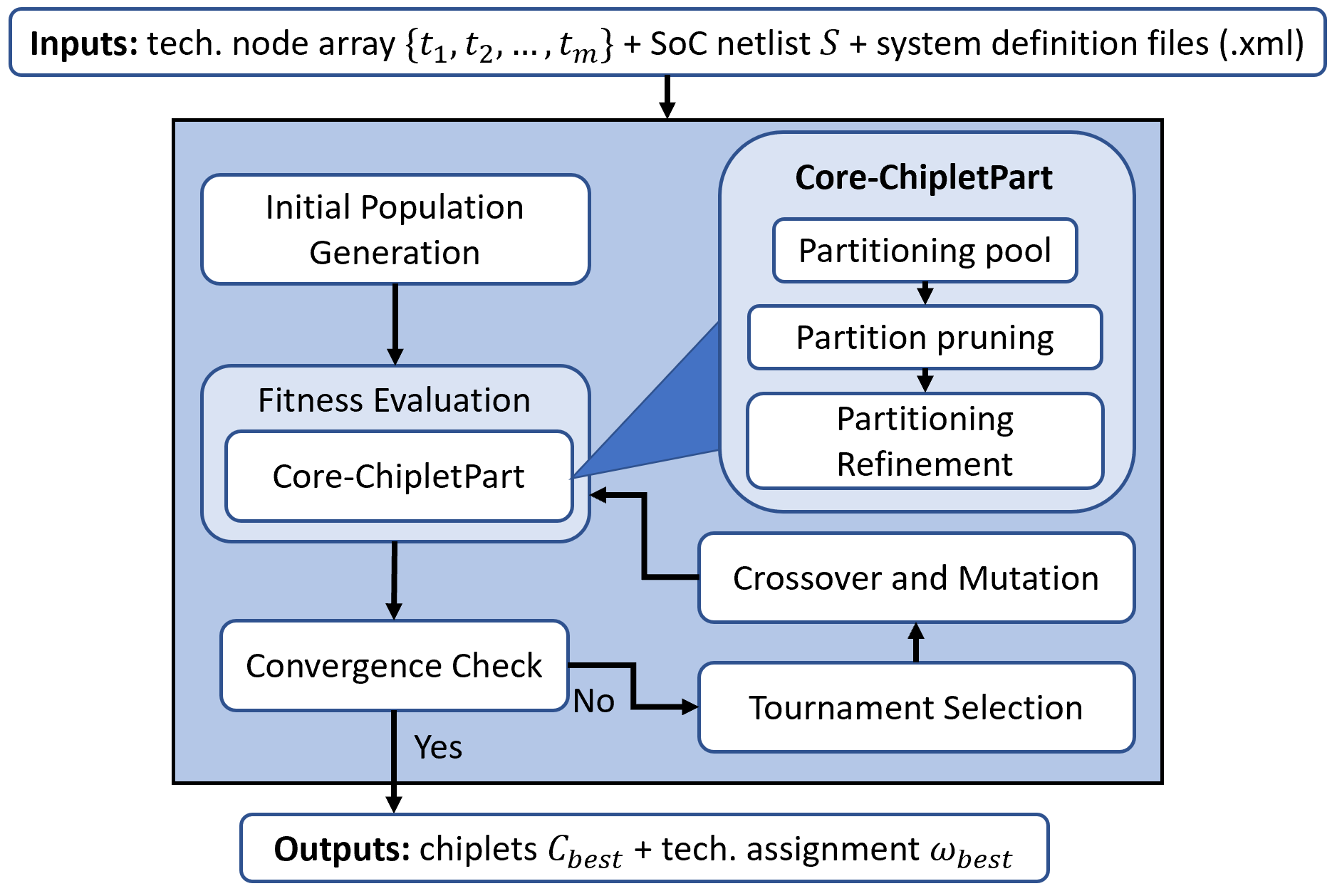}
    \Description{A flowchart diagram showing the inputs and outputs of the ChipletPart framework.}
    \caption{\textcolor{black}{{\em ChipletPart} Framework. {\em Core-ChipletPart} is shown in Figure~\ref{fig:core_chipletpart}. Partitioning refinement is shown in Figure~\ref{fig:cost_fm}.}}
    \label{fig:overall_chipletpart}
\end{figure}

\subsection{Overview of {\em ChipletPart} Framework}
\label{sec:chipletpart_flow}

We now describe our chiplet partitioning framework, 
\textit{ChipletPart} (see Figure~\ref{fig:overall_chipletpart}). 
Our \textit{ChipletPart} differs from classical min-cut 
partitioners~\cite{KarypisAKS99,bustany2022specpart,BustanyGKKPW23} 
and previous 
approaches~\cite{Sangiovanni-VincentelliLZZ23,Li24, ChenLZZL23, Kabir20} 
in several key respects. 
(i) We optimize a comprehensive chiplet cost function 
(Section~\ref{sec:cost_driver}), unlike min-cut partitioners 
that focus on minimizing net cutsize.
(ii) We integrate a fast, \textcolor{black}{Go-With-the-Winners}~\cite{Aldous94} SA-based chiplet floorplanner to 
ensure I/O-feasibility of \textcolor{black}{a} partitioning 
solution (Section~\ref{sec:floorplan}).
(iii) We leverage a \textcolor{black}{genetic algorithm} (GA) framework to discover high-quality 
technology node assignments for heterogeneous integration.

\begin{figure}
    \centering
    \includegraphics[width=0.8\columnwidth]{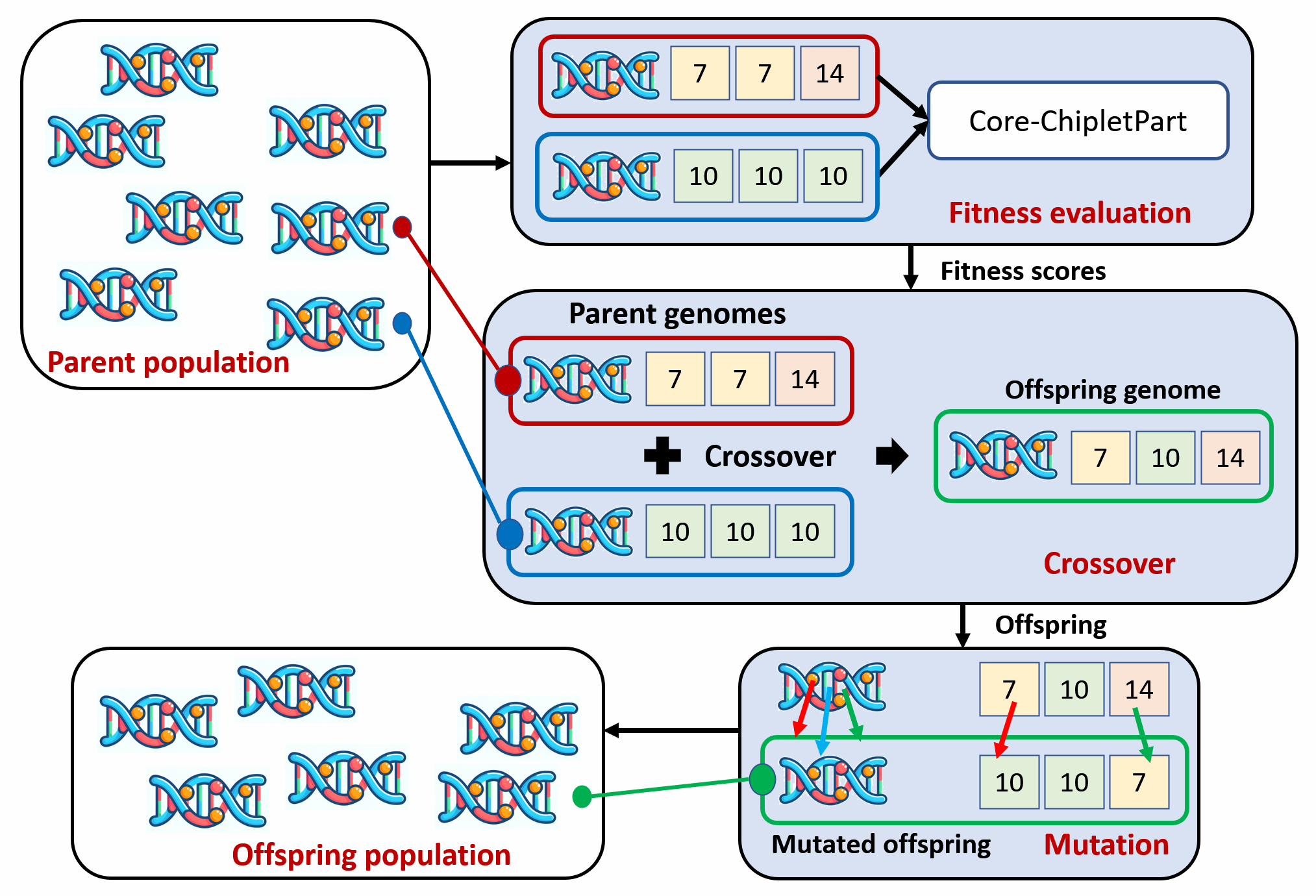}
    \Description{A diagram showing the behavior of the genetic algorithm including crossover and mutation.}
    \caption{\textcolor{black}{GA-based technology assignment.
A genome is a set of technology nodes. Shown: 
two parent genomes $\langle 7, 7, 14 
\rangle$ and $\langle 10, \textcolor{black}{10}, 10 \rangle$
undergo crossover to produce the offspring
$\langle 7, 10, 14 \rangle$; the
offspring undergoes mutation to produce 
$\langle 10, 10, 7 \rangle$ for the 
next-generation population.}}
    \label{fig:genetic_illus}
\end{figure}

{\em ChipletPart} inputs include a system block-level 
netlist $\mathcal{S}$, a set of technology nodes $\mathcal{T}$, 
and system definition provided in XML
files.\footnote{As in \cite{GraeningPG23, Graening_CATCH}, our netlists 
are modeled as graphs rather than hypergraphs, \textcolor{black}{since 
these previous works use a
directed (source-sink) edge for every inter-block connection.}}
\textcolor{black}{The algorithm is outlined in 
Algorithm~\ref{alg:overall_chipletpart} and visually illustrated 
in Figure~\ref{fig:genetic_illus}. 
Within the GA framework, a {\em gene} is a technology node,
and a {\em genome} is a sequence of technology nodes in which
corresponding chiplets (partitions of $\mathcal{S}$) will 
be implemented.
Details and source code for our GA-based partitioning and technology 
assignment methodology are in \cite{anonymousrepo}.}

\noindent
\textbf{Step 1: \textcolor{black}{[Lines 3-4]}} We first generate the 
initial population in the GA.
\textcolor{black}{Each member in the population (a {\em genome}) 
represents an ordered sequence of technology assignments (to chiplets) 
$\omega: \mathcal{C} \to \mathcal{T}$.}
We generate $tot_{pop}$ ($tot_{pop}$ = 50 by default) genomes; 
each genome is generated 
by randomly mapping $K_{\text{max}}$ chiplets 
to technology nodes ${\mathcal{T}}$.\footnote{For simplicity, we set $tot_{pop} = 
    k_{pop} + \sigma$.} 
\textcolor{black}{To avoid redundant evaluations and improve scalability, we 
canonicalize all generated genomes by mapping functionally equivalent assignments to a unique representation. For example, genomes \textcolor{black}{$\langle 7\,\text{nm}, 7\,\text{nm}, 14\,\text{nm} \rangle$ and $\langle 14\,\text{nm}, 7\,\text{nm}, 7\,\text{nm} \rangle$} are treated as equivalent, 
since they correspond to the same multiset of assignments.}

\noindent
\textbf{Step 2: \textcolor{black}{[Lines 5-11]}}  
We assess the fitness of each genome in the current population. 
For each genome $\omega_j$, 
we (i) run \textit{Core-ChipletPart}\footnote{For better
scalability, we run {\em Core-ChipletPart} with 
fewer FM (Fiduccia-Mattheyses) moves and passes, and 
fewer \textcolor{black}{{\em initial}} solutions. Details are seen in our code 
\cite{anonymousrepo}.} to generate a partitioning 
solution based on $\omega_j$
and (ii) calculate the {\em fitness score} of this solution using our 
cost model.

\noindent
\textcolor{black}{\textbf{Step 3: \textcolor{black}{[Lines 12-21]}}
We declare convergence if either of the following two
termination criteria is met:}
\begin{itemize}
    \item \textcolor{black}{The algorithm reaches the maximum
    number of generations, $\Psi$ ($\Psi = 50$ by default). Upon
    reaching this limit, the algorithm terminates and returns the
    best chiplet solution, $\mathcal{C}_{best}$, along with its
    corresponding technology mapping, $\omega_{best}$.}
    
    \item \textcolor{black}{The improvement between successive 
    generations falls below a predefined threshold,
    $\Delta_{threshold}$ ($\Delta_{threshold} = 0.01$ by default),
    and remains below this threshold for more than $\epsilon$
    consecutive generations ($\epsilon = 10$ by default). If this
    condition is met, the algorithm terminates and returns $\mathcal{C}_{best}$ and $\omega_{best}$.}
\end{itemize}

\noindent
\textcolor{black}{\textbf{Step 4: \textcolor{black}{[Line 22]}} 
We use {\em tournament} selection~\cite{miller1995genetic} 
to identify 
candidate parents for crossover. Rather than evaluating the
entire population at once, the selection process operates on randomly 
chosen
``tournaments'', each consisting of $\zeta$ genomes competing based on 
their fitness scores. The genome with the 
highest fitness in each tournament is
selected as a candidate parent. This procedure is repeated 
until $k_{pop}$
pairs of parents are chosen.\footnote{\textcolor{black}{A genome can be selected 
multiple times as a parent.}}}

\noindent
\textcolor{black}{\textbf{Step 5: [Lines 23-25]} 
We generate the next-generation population (\textcolor{black}{offspring}) 
through crossover, elitism, and mutation.
Notably, we use uniform crossover \cite{syswerda1989uniform} and random 
resetting mutation ~\cite{mitchell1998introduction}. 
Additionally, we retain the top $\sigma$ genomes from the parent 
generation based on their fitness scores (elitism), 
ensuring that high-quality solutions persist across generations.} 

In our implementation, we set $tot_{pop} = 50$, $\zeta = 3$, $k_{pop} = 45$, $\sigma = 5$ and $K_{max} = 8$ (Section~\ref{subsubsec:hyperparam_explore}). So, we conduct $90$ tournaments (each tournament picks \textcolor{black}{three} genomes from the current generation) to select $90$ parent genomes, which are then paired to generate $45$ offspring genomes through crossover and mutation. We then supplement these $45$ offspring with the top \textcolor{black}{five} genomes from the current generation (elitism) \textcolor{black}{to} generate the next generation population of $50$ genomes. For additional implementation details, we refer the reader to \cite{anonymousrepo}.


\noindent
\textbf{Choice of optimizers:} Our choice of picking GA and SA is motivated by their ability to flexibly 
optimize over {\em black-box}, non-convex and
highly discontinuous cost landscapes --- characteristics inherent to our chiplet cost model (Section~\ref{sec:cost}). In contrast, ILP-based methods are less suitable for our framework due to the lack of closed-form or linearizable expressions for I/O-feasibility and chiplet 
cost functions.
GA enables direct 
co-optimization of technology assignment and chiplet partitioning, while SA ensures that floorplanning solutions are always I/O-feasible. \textcolor{black}{Having a unified optimization approach and formulation} is challenging since the solution space spans both discrete combinatorial variables 
(chiplet-to-tech assignment) and continuous geometric constraints
(\textcolor{black}{I/O} reach), which are difficult to encode and optimize jointly. 
We hence opt for a modular decomposition --- using GA for assignment and SA for feasibility evaluation.
\textcolor{black}{Note that Bayesian optimization (BO) is another} powerful
tool for optimizing complicated, black-box functions --- we explore its potential
as an alternative optimizer in Section~\ref{subsec:bayesian_opt}.

In the following sections, we discuss the chiplet cost model, 
\emph{Core-ChipletPart}, 
and the SA-based floorplanner.

\begin{algorithm}[!t]
    \footnotesize
    \SetKwInput{Input}{Input}
    \SetKwInOut{Output}{Output}
    \Input{Standard Inputs: $\mathcal{T}$, $\mathcal{S}$ \\
           \quad \quad \quad Hyperparameters: $tot_{pop}$, $k_{pop}$, $\Delta_{threshold}$, \textcolor{black}{$\epsilon$,
           $\zeta$, $\Psi$, $\sigma$}}
    \KwOut{Chiplet partitioning solution $C_{best}$, \\  
        \quad \quad \quad   Technology assignment $\omega_{best}$}

    \SetAlgoNoEnd    
    \BlankLine
    \textcolor{black}{$\Delta_{threshold} \leftarrow 0.01$; 
    $\epsilon \leftarrow 10$ ;  
    $\zeta \leftarrow 3$;
    $\Psi \leftarrow 50$;  
    $\sigma \leftarrow 5$;}  \\
    $tot_{pop} \leftarrow 50$; $k_{pop} \leftarrow 45$; $p_c \leftarrow \textcolor{black}{0.60}$; $p_m \leftarrow \textcolor{black}{0.07}$;  \textcolor{black}{$K_{\text{max}} \leftarrow 8$;}\\
    
    \tcc{\textcolor{brown}{1. Initial population generation}}    
    $Cost_{best} \leftarrow \infty$; \quad $Cost_{prev} \leftarrow \infty$; \quad $\Delta \leftarrow \infty$;  \\     
    Generate the initial population with $tot_{pop}$ genomes
    where each genome is a random mapping from $K_{\text{max}}$ chiplets to ${\mathcal{T}}$; \\   

     \textcolor{black}{$i \leftarrow 0;$ $num\_iters \leftarrow 0$}\textcolor{black}{;} \\ 
    \While{true} {
        \tcc{\textcolor{brown}{2. Fitness evaluation}}
        \ForEach{$\omega_j$ in the current population}{
            $\mathcal{C}_{\omega_j} \leftarrow$ Generate the partitioning solution using \textit{Core-ChipletPart} with 
            technology assignment $\omega_j$; \\
            $Cost_{\omega_j} \leftarrow$ Calculate the cost of $\mathcal{C}_{\omega_j}$ (Section \ref{sec:cost_driver}); \\
            \If{$Cost_{\omega_j} \leq Cost_{best}$}{
                $Cost_{best} \leftarrow Cost_{\omega_j}$;  \quad
                $\mathcal{C}_{best} \leftarrow \mathcal{C}_{\omega_j}$; \quad
                $\omega_{best} \leftarrow \omega_j$; \\
            }
        }

        \tcc{\textcolor{brown}{3. Convergence check}}
        \textcolor{black}{\If{$num\_iters \geq \Psi$} {
            \KwRet{$C_{best}, \omega_{best}$}
        }}
        $\Delta \leftarrow$ $Cost_{prev} - Cost_{best}$; \\  
        \textcolor{black}{\If{$\Delta \leq \Delta_{threshold}$} {
            $i \leftarrow i + 1$; \\ 
            \If{$i > \epsilon$} {
                \KwRet{$C_{best}, \omega_{best}$}
            }
        } \Else {
            $i \leftarrow 0$; \\ 
        }}
        $Cost_{prev} \leftarrow Cost_{best}$; \\

        \tcc{\textcolor{brown}{4. Tournament selection}}

        \textcolor{black}{Select $k_{pop}$ pairs of parents 
        using tournament selection with tournament size $\zeta$;} \\

        \tcc{\textcolor{brown}{5. Crossover, elitism and mutation}}
        Perform crossover on each selected pair $(\omega^p_x, \omega^p_y)$ to generate offspring $\omega^o_x$, 
        using a crossover probability $p_c$; \\
        \textcolor{black}{Apply mutation to each offspring $\omega^o_x$ with mutation probability $p_m$;} \\ 
        \textcolor{black}{Construct the next generation by selecting the top $\sigma$ elite genomes from the current generation, 
        along with $k_{pop}$ \textcolor{black}{offspring};} \\
    }
    \caption{\textcolor{black}{Overall \textit{ChipletPart} framework.}}
    \label{alg:overall_chipletpart}
\end{algorithm}

\section{Chiplet Partitioning Drivers}
\label{sec:cost_driver}

\textcolor{black}{Smaller chiplets can potentially \textcolor{black}{bring} lower costs 
and improved yield,
while technology heterogeneity offers better power and performance. The
overhead of inter-chiplet \textcolor{black}{I/O} also affects how a system 
should be partitioned
into chiplets. In this section, we briefly discuss these factors.}

\subsection{Chiplet Cost Model} 
\label{sec:cost}

We use the open-source cost model from 
\cite{Graening_CATCH,costmodelUCLA}. It computes cost and 
yield based on the 
chiplet/interposer area (dependent on the technology node), assembly 
processes (dependent on bonding parameters), 
and \textcolor{black}{I/O} placement constraints (dictated by netlist connectivity and 
reach). The model calculates the individual cost and 
yield of each chiplet along with the interposer, and then aggregates 
them for the full assembly cost. We consider designs 
\textcolor{black}{to be} high volume in \textcolor{black}{our studies}, to minimize the impact of NRE. The cost 
model is summarized by Equation~\ref{eq:part_cost}. For further details, 
see \cite{Graening_CATCH}. \textcolor{black}{Benefits of using the cost model in~\cite{Graening_CATCH} are discussed in Section~\ref{subsec:validations_of_par} with results in Tables~\ref{tab:partitioning_costs_only}
and~\ref{tab:partitioning_feas_only}.}

A block will have different areas and power in different 
technology nodes. \textcolor{black}{While assigning a block to a more advanced technology node could, 
in principle, improve performance, we assume a fixed system architecture in this work and therefore
\textcolor{black}{keep} performance constant while scaling power accordingly.\footnote{\textcolor{black}{Modeling architectural changes based on
block-to-technology assignments is beyond the scope of the 
    current work and is left as a future direction.}}}
\textcolor{black}{In this work,} we follow \cite{Stilmaker2017} to model this scaling.
Different technology nodes result in different cost per unit area 
for chiplets, 
along with differences in yield and reticle-fit dependent lithography 
costs \cite{Graening_CATCH}. We also scale memory and logic with 
different factors \cite{heyman_2024} since memory 
tends to scale poorly for advanced nodes.
Different technology nodes will also have 
different non-recurring design \cite{Lapedus2018} and 
manufacturing costs \cite{patel_dark_2023}.  For example, the transition 
between \textcolor{black}{45 $\text{nm}$} and 
\textcolor{black}{10 $\text{nm}$} reduces logic size by a factor of  $\sim$10$\times$ 
\cite{Stilmaker2017}, while NRE design cost increases by 
$\sim$4.6$\times$ \cite{Lapedus2018} and general cost per wafer 
increases by $\sim$2.6$\times$ \cite{Shilov2020}.

\subsection{Chiplet Power Model}
\label{sec:power_model}
We assume that performance of chiplets is preserved 
across technology nodes but that the power changes. We further assume 
that inter-chiplet communication latency is small enough to 
not influence 
the architecture. If certain inter-block interfaces are highly
latency-sensitive, then those blocks should be merged in our 
framework \textcolor{black}{to} 
map them to the same chiplet. More complex performance models as 
in  \cite{ChenLZZL23}  are possible but not explored in this work. 

Power is calculated as the sum of block power and \textcolor{black}{I/O} power. 
 If all blocks are in 
the same chiplet, there will be no additional power due to \textcolor{black}{I/O}. 
However, if the blocks are placed into multiple partitions (chiplets), some 
additional power will be consumed by the added \textcolor{black}{I/O} cells. 
We scale the block power according to the scaling factors 
in \cite{Stilmaker2017} for different technology nodes.\par

\subsection{\textcolor{black}{I/O} Reach Model}
\label{subsec:io_reach_model}
We use an \textcolor{black}{I/O} cell model where each \textcolor{black}{I/O} cell has both an area and 
a ``reach'' which is the maximum wire length that can be driven 
by the \textcolor{black}{I/O} cell. The reach depends on the \textcolor{black}{I/O} transceiver design and is 
often dictated by the inter-chiplet \textcolor{black}{I/O} standard. \textcolor{black}{For example}, 
\textcolor{black}{{\em Universal Chiplet Interconnect Express}} (UCIe) \cite{UCIe1p0} 
guarantees a \textcolor{black}{reach} of \textcolor{black}{2 $\text{mm}$} for advanced packages while 
standard packages (such as organic substrates) 
support larger reaches of up to \textcolor{black}{25 $\text{mm}$}.

\textcolor{black}{Since we assume a fixed system architecture, our partitioner does not modify the
network structure or insert pipeline stages to compensate for long inter-chiplet wires. We also assume a standardized \textcolor{black}{I/O} cell with a fixed maximum reach instead of more complicated \textcolor{black}{I/O} structures. 
These assumptions help keep the partitioning problem well-scoped and tractable. We leave sophisticated \textcolor{black}{I/O} planning including pass-through connections, inserting buffers on the substrate, multiple \textcolor{black}{I/O} types per design, etc. to future work. 
However, users can explore multiple architectural variants and \textcolor{black}{I/O} types for the same design. 
For example, in our industry testcase (Section~\ref{subsec:benchmarks}), we evaluate two versions that differ only 
in their crossbar configurations \textcolor{black}{ --- }and in Figure \ref{fig:case_studies}, we \textcolor{black}{examine} the impact of different \textcolor{black}{I/O} cell types on partitioning.}

\begin{figure}
    \centering
    \includegraphics[width=0.55\columnwidth]{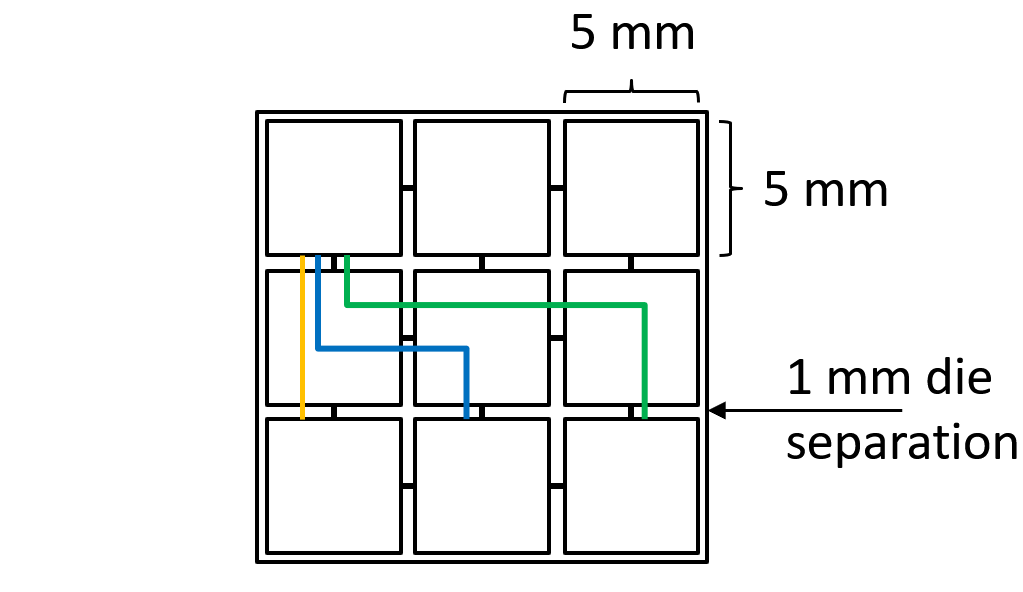}
    \Description{An image showing squares in a 3x3 configuration representing chiplets with connections shown as colored lines of various lengths to indicate the different reach requirements fore connections from one chiplet to another in the grid.}
    \caption{Illustration of {\em reach}. \textcolor{black}{Wirelengths: black - \textcolor{black}{1 $\text{mm}$, orange - 7 $\text{mm}$, blue - 13 $\text{mm}$, and green - 19 $\text{mm}$}.}}
    \label{fig:reach_example}
\end{figure}

Consider the example in Figure \ref{fig:reach_example}, 
and assume that a bundle of wires 
is connected from the side of one chiplet to the side of 
another. 
If the chiplets are \textcolor{black}{5 $\text{mm}$} on a side and spaced apart 
by \textcolor{black}{1 $\text{mm}$}, then the short 
black connections require a reach of at least $1~\text{mm}$, 
while the \textcolor{black}{orange, blue, and green connections require reaches of at least 7 $\text{mm}$, 13 $\text{mm}$, and 19 $\text{mm}$ respectively}. 
Small values of reach constrain the floorplan, while 
larger values of reach 
come at the cost of more expensive \textcolor{black}{I/O} cells.\par

\textcolor{black}{I/Os} are  discretized depending on the standard. For instance, a single 
UCIe \textcolor{black}{I/O} cell is as large as \textcolor{black}{0.88 $\text{mm}^2$} \cite{UCIe1p0} while for 
parallel signaling (e.g., \textcolor{black}{Advanced Interface Bus (AIB)} or \cite{Pal2021}), an \textcolor{black}{I/O} cell can be as 
small as \textcolor{black}{157 $\text{$\mu$m}^2$} \cite{PalIO2021}. This discretization can \textcolor{black}{add} 
complexity for chiplet partitioning, as the \textcolor{black}{I/O} (i.e., net cut) cost 
calculation follows a \textcolor{black}{stepwise} curve as a function of net cut. 

\section{Chiplet Partitioning Core}
\label{sec:partition}
In this section, we first discuss our chiplet-cost-driven partitioning 
approach \textcolor{black}{{\em Core-ChipletPart}} in Section \ref{sec:partition_refine}.
Then, we present our reach-aware chiplet floorplanning approach in
Section \ref{sec:floorplan}.

\subsection{Cost-driven Partitioning: \textcolor{black}{{\em Core-ChipletPart}}}
\label{sec:partition_refine}

\begin{figure}
    \centering
    \includegraphics[width=0.5\columnwidth]{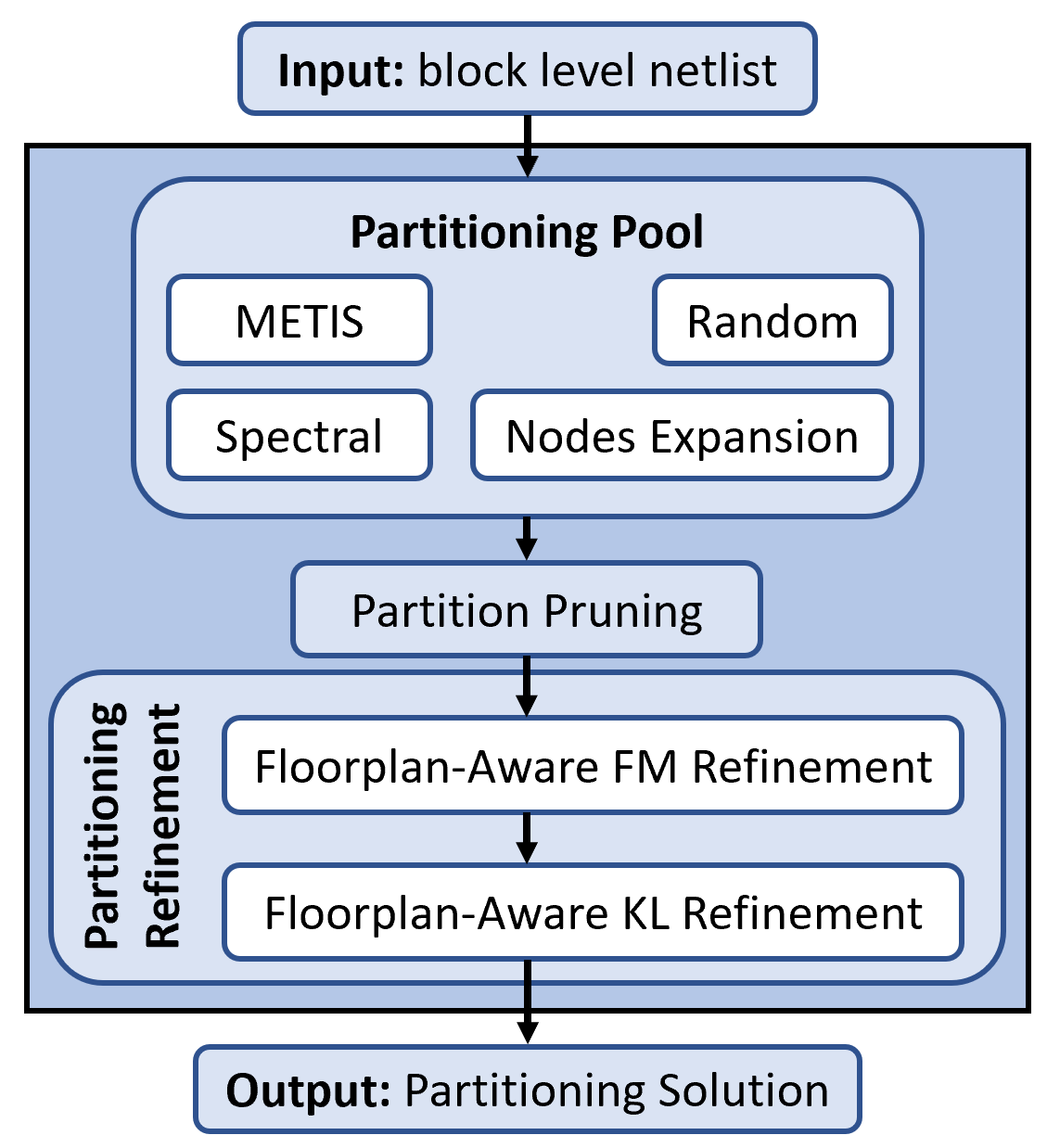}
    \Description{A flowchart showing the Core-ChipletPart partitioning flow.}
    \caption{{\em Core-ChipletPart} partitioning
    flow. We use multiple \textcolor{black}{techniques} to generate the initial partitions, then we prune out the worst-performing initializations before running refinement.}
    \label{fig:core_chipletpart}
\end{figure}

Our {\em Core-ChipletPart} does not adopt the widely-used 
multilevel approach
since there are typically only a few hundreds of IP blocks 
in a block-level netlist.
As shown in Figure~\ref{fig:core_chipletpart}, we 
(i) compute a large pool of initial 
partitioning solutions using 
a variety of graph partitioning algorithms; (ii) prune 
poor-quality partitioning solutions 
from the pool; and (iii) perform floorplan-aware Fiduccia–Mattheyses\textcolor{black}{~\cite{1585498}}
(FM)-based and \textcolor{black}{Kernighan-Lin\textcolor{black}{~\cite{HELSGAUN2000106}} (KL)}\textcolor{black}{-based} refinement. We then 
output the solution with the best cost.

\label{sec:init_part}

\noindent
{\bf Generation of a partitioning pool.}
Similar to the \textcolor{black}{{\em initial partitioning}} stage in \textcolor{black}{widely-used} multilevel partitioning approaches 
\textcolor{black}{\cite{KarypisAKS99, BustanyGKKPW23}}, 
we compute a pool of initial partitioning solutions so as to explore 
a larger solution space.\footnote{However, unlike 
multilevel approaches \textcolor{black}{\cite{KarypisAKS99, BustanyGKKPW23}},
{\em Core-ChipletPart} omits the coarsening stage due to 
the relatively small number of vertices (i.e., 384 IP blocks 
in our largest testcase, as shown in Table \ref{tab:testcases}) \textcolor{black}{in the netlist graph}.}
Specifically, we use the following graph partitioning 
methods.

\begin{itemize}[noitemsep,topsep=0pt,leftmargin=*]
    \item \textit{Spectral partitioning:} \textcolor{black}{Because the 
    block-level netlist is both small and sparse, we apply 
    \emph{spectral partitioning}~\cite{von2007tutorial}. 
We first compute the spectral embedding using the two smallest 
nontrivial eigenvectors, then 
cluster the IP blocks with the K-means 
algorithm~\cite{likas2003global}.\footnote{
In our implementation, we use a parallel K-means method with smart 
initialization (K-means++~\cite{bahmani2012scalable}), 
and we set $K=4$ by default.}} 
    \item \textit{\textcolor{black}{High-degree} nodes expansion:} \textcolor{black}{Our 
    node expansion approach follows \cite{zhang2017graph}.}
    In the block-level netlist, high-degree nodes (e.g., crossbars) are 
    distributed across different chiplets, 
    and then expanded via breadth-first search. If multiple chiplets 
    qualify for expansion, 
    the block is assigned to the chiplet with which it has the strongest
    connection.
    \item \textit{Random partitioning:} We randomly distribute all 
    blocks across different chiplets. 
    \textcolor{black}{We generate multiple random solutions while 
    sweeping $K = \{1, .. , K_{\max}\}$.\footnote{Our studies show that $K_{\max}=8$ is sufficient for 
    our testcases. If the solution finds $K_{\max}$ number of partitions, the user can increase $K_{\max}$ to check larger 
    numbers of partitions.} We consider $K = 1$ to 
    allow the degenerate case where
    all the blocks are placed in a single chiplet.}
    \item \textit{METIS:} \textcolor{black}{We use the METIS graph 
    partitioner~\cite{KarypisK96} to create initial partitioning 
    solutions, 
    using the METIS APIs from~\cite{metis_repo}. Similar to our random 
    partitioning approach, we sweep $K = \{2, .. , \textcolor{black}{K_{max}} 
  \}$.}
\end{itemize}

Our partitioning pool consists of 11 solutions in total: one from the spectral method, one from node expansion, five from random partitioning, and four from METIS. We next discuss the pruning step, which discards low-quality solutions from this pool.


\noindent
{\bf Partitioning solution pruning.}
\textcolor{black}{We use a statistical filtering mechanism 
to prune the partitioning pool. First, we compute the cost of each initial
partition using our cost model (Section~\ref{sec:cost}), 
and then calculate
the mean and standard deviation of these costs, 
as well as each partition’s
cost relative to the best solution. Our pruning applies two complementary
thresholds: (i) a Z-score~\cite{abdi2007z} threshold, eliminating 
partitions whose costs exceed 1.5 standard deviations 
above the mean and (ii) a relative cost threshold, 
removing partitions whose costs are worse than twice the minimum cost.
To prevent over-pruning, we retain at least three solutions 
even if they all
appear statistically poor. Pruning significantly reduces 
computational workload by
eliminating low-quality candidates early.}

\begin{figure}
    \centering
    \includegraphics[width=1.0\columnwidth]{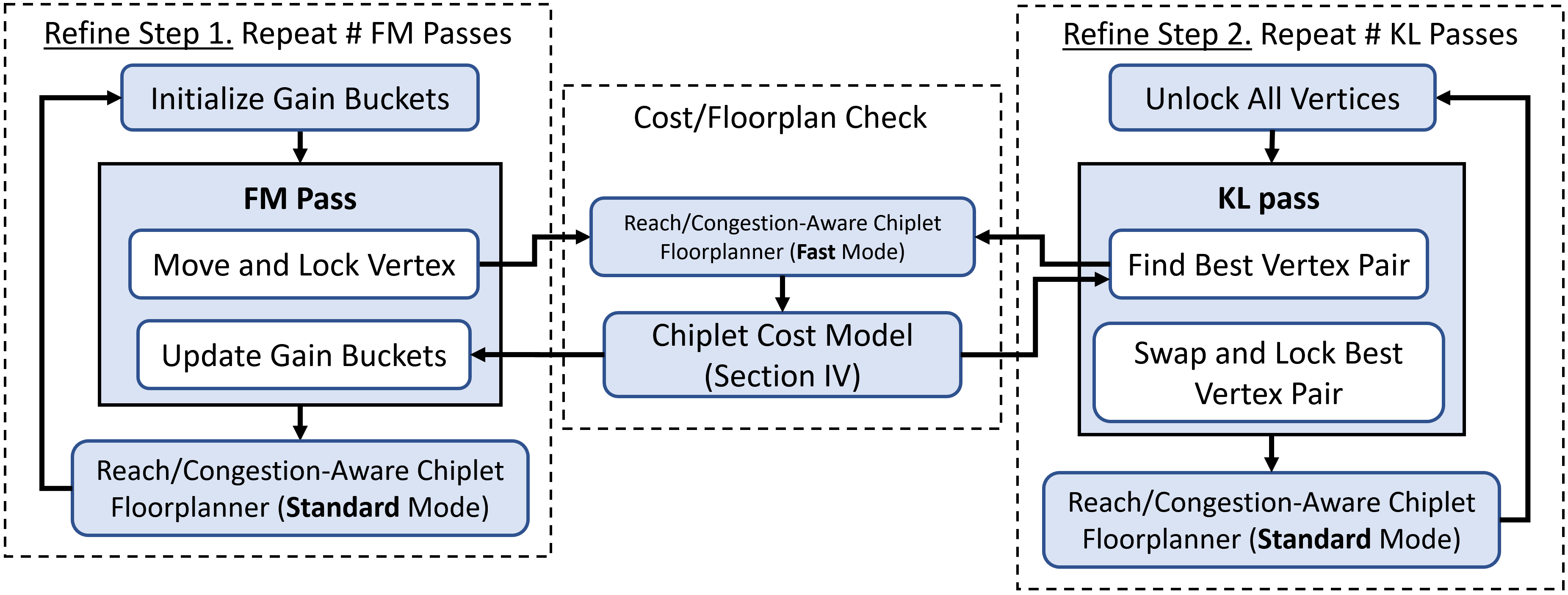}
    \Description{An flowchart showing the behavior of partition refinement.}
    \caption{Floorplan-aware FM and KL-based refinement. In our implementation, \textcolor{black}{we run FM first, followed by KL}.}
    \label{fig:cost_fm}
\vspace{-0.1in}
\end{figure}

\noindent 
\textcolor{black}{\bf Partitioning refinement.}
\textcolor{black}{We use \textcolor{black}{two strategies for refinement}: (i) FM and (ii) KL. 
Both \textcolor{black}{of} these processes are shown in Figure~\ref{fig:cost_fm}. 
Each potential FM move or \textcolor{black}{KL swap} triggers a call to the chiplet cost model to obtain
the cost (gain) of an updated partitioning solution. This evaluation includes generating a floorplan 
solution using the ``fast'' mode of our SA-based 
reach-aware chiplet floorplanner, followed by calculating the 
corresponding cost. After completing each FM or KL pass, the 
``standard'' mode of the SA-based reach-aware chiplet floorplanner is 
employed to search for an improved floorplan solution 
for the subsequent pass. Further details about our SA-based reach-aware 
chiplet floorplanner are provided in Section \ref{sec:floorplan}.}
\textcolor{black}{We adopt the K-way FM implementation from {\em TritonPart}~\cite{BustanyGKKPW23}, 
while our KL refinement\textcolor{black}{~\cite{klrefine}} follows the methodology described in~\cite{HELSGAUN2000106}. In our implementation, we apply KL 
after FM to further improve solution quality, as KL explores a broader neighborhood by enabling
pairwise vertex swaps --- compared to FM’s single-vertex moves.}

\subsection{Reach-aware Chiplet Floorplanning}
\label{sec:floorplan}

\begin{figure}[!htb]
    \centering
    \includegraphics[width=0.7\columnwidth]{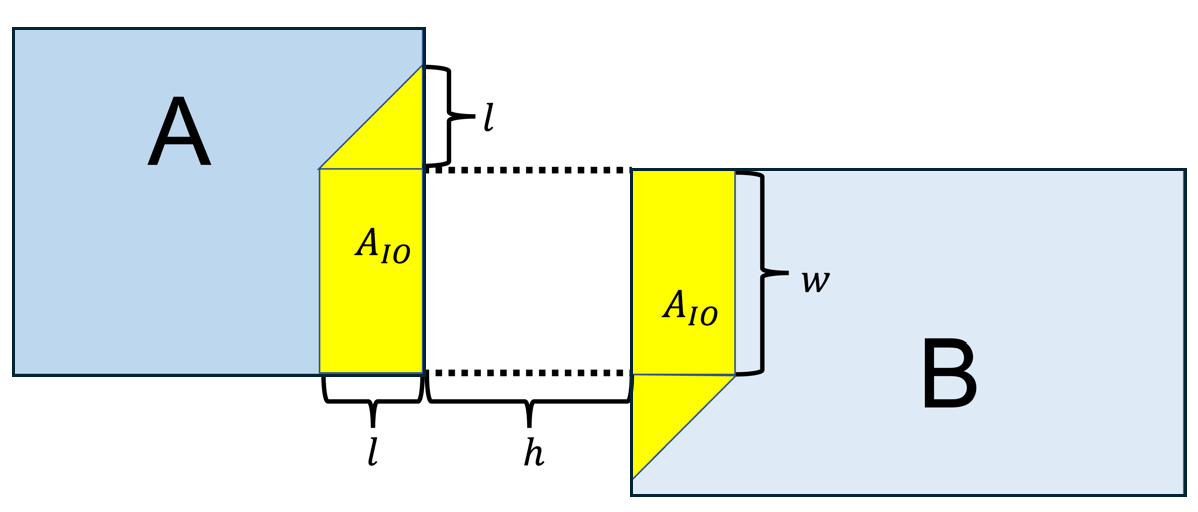}
    \Description{Diagram showing two chiplets (A and B) represented by rectangles placed next to each other with distances shown in the corresponding equation marked.}
    \caption{Illustration of wirelength computation ($length(e)$).
    \textcolor{black}{The yellow region indicates the placement 
    area for \textcolor{black}{I/O} cells that \textcolor{black}{will satisfy} the reach constraint.}}
    \label{fig:length}
\end{figure}

Chiplet floorplanning is a crucial step, 
as it determines the location and shape of each chiplet on 
the interposer.
In contrast to the well-studied macro placement, 
chiplet floorplanning must also consider {\em I/O-feasibility}.
More specifically, the wirelength of a net must not exceed the reach 
specified by the corresponding \textcolor{black}{I/O} cell (Section \ref{sec:cost_driver}).
Figure \ref{fig:length} illustrates the wirelength (\textcolor{black}
{measured in terms of Manhattan distance}) calculation for net 
$e$ connecting chiplets $A$ and $B$. 
Assuming that the \textcolor{black}{{\em bitwidth}} of net $e$ is $n_e$ and the area of \textcolor{black}{I/O} cells 
is $A_{IO}$, 
then the wirelength of net $e$ ($length(e)$) 
\textcolor{black}{in the figure}
is calculated as: (i) $ l = \sqrt{w^2 + 2 A_{IO}} - w$ and
(ii) $length(e) = h + 2l$,
where $l$ is the \textcolor{black}{{\em depth}} needed for all the \textcolor{black}{I/O} 
cells.\footnote{\textcolor{black}{The detailed code for more general 
cases is available in \cite{reach}}.}
Then the reach violation \textcolor{black}{penalty} for 
net $e$ is $n_e \times max(length(e) - reach(e), 0.0)$.
The reach violation \textcolor{black}{penalty} for a chiplet is defined as 
the summation of the reach violation
\textcolor{black}{penalties} for all the nets connected to the chiplet.


\noindent
{\bf Problem formulation.}
Our work uses a novel reach-aware chiplet floorplanning approach.
The input to our floorplanner is a set of chiplets $ \mathcal{C} $,
and a chiplet-level netlist $ \mathcal{N} $ that defines 
the connections between chiplets.
The output is a floorplanning solution which provides the location 
and shape (width and height) of each chiplet.
The objective of our floorplanner is
\begin{equation}
   \label{eq:reach_cost}
    WL_{reach} = \sum_{e \in N}{n_e \times max(length(e) - reach(e), 0.0)}
\end{equation}
\begin{equation}
\label{eq:objective_sa}
    \text{\textbf{min}} \quad \alpha \times WL_{reach} + \beta \times A_{C} + \gamma \times A_{P} 
\end{equation}
where \textcolor{black}{$WL_{reach}$, $A_{C}$, and  $A_{P}$} \textcolor{black}{respectively denote} \textcolor{black}{the reach 
violation penalty, the area of chiplets,
and the area of the package.}
$\alpha$, $\beta$ and $\gamma$ are user-defined 
coefficients that can be modified
to achieve a desired trade-off between \textcolor{black}{the different objectives in Equation~\ref{eq:objective_sa}}.\footnote{The 
default values for $\alpha$, $\beta$ and $\gamma$ 
are $1.0$, $1.0$ and $1.0$, respectively.}
Note that we allow the area of chiplets to increase to satisfy 
the reach constraints.
During optimization, the following constraints are considered:
\begin{itemize}[noitemsep,topsep=0pt,leftmargin=*]
    \item \textbf{Overlap constraint}: No two chiplets can overlap.
    \item \textbf{Separation constraint}: In practical chiplet-based systems, 
    extra space must exist between neighboring chiplets 
    \textcolor{black}{to account for dicing and alignment accuracy,
    and to prevent defects and mechanical stress during chiplet assembly.}
    The separation constraint sets the minimum distance between any two chiplets.
\end{itemize}

\begin{figure}[!t]
    \centering
    \begin{subfigure}[b]{0.75\textwidth}
        \centering
        \includegraphics[width=1.0\textwidth]{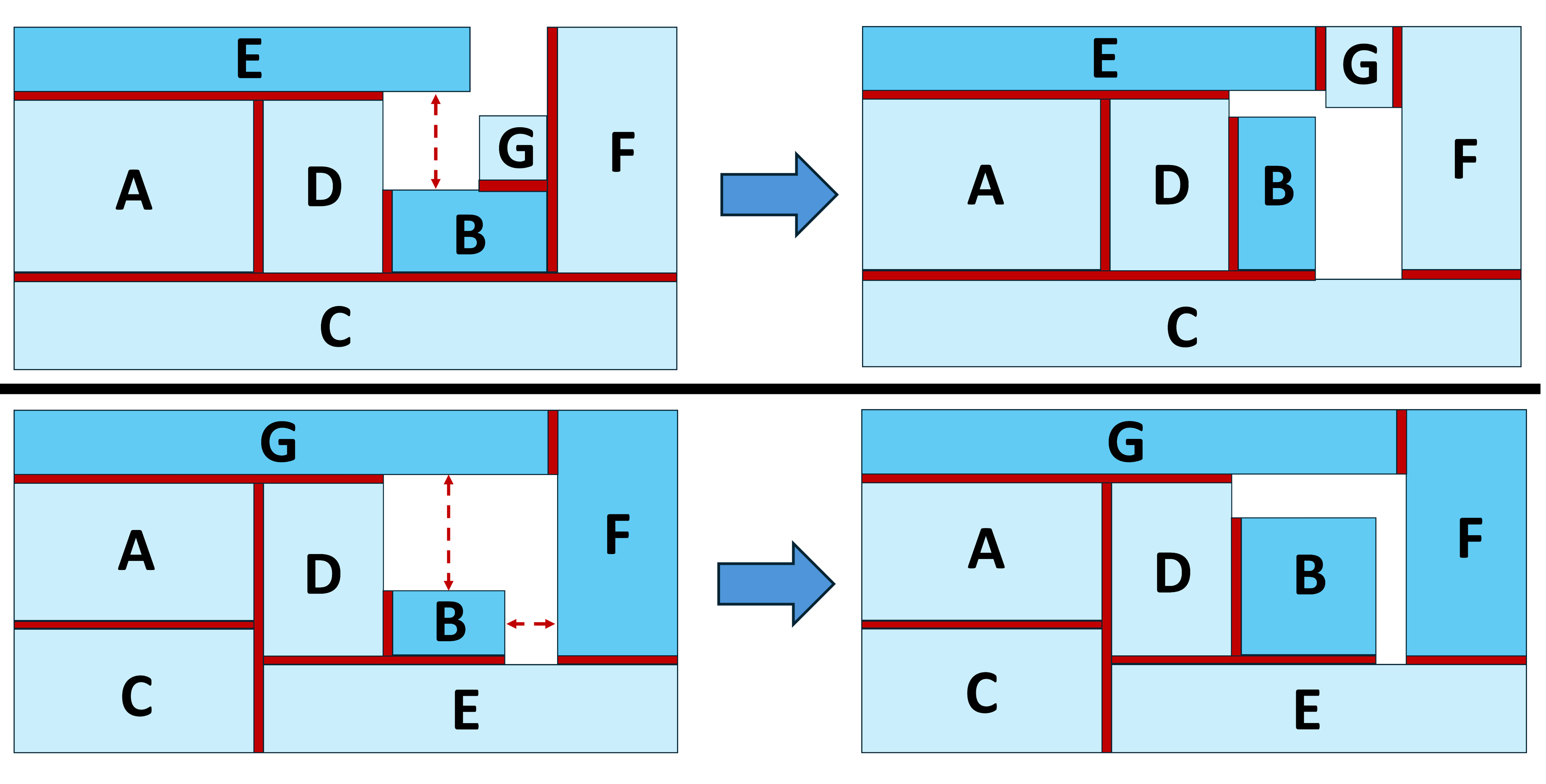}
    \end{subfigure}
    \Description{Image showing different possibilities for resizing chiplets.}
    \caption{Top: Resizing of chiplet B fixes a reach violation for a net connecting chiplets B-E. 
    Bottom: Expansion (bloating) of chiplet B fixes reach violations 
    for nets connecting chiplets B-G
    and B-F. In each figure pair, left = before; right = after. 
    Red regions indicate the separation 
    constraint between chiplets, and red dashed arrows indicate 
    reach violations before fixing.}
    \label{fig:resize-expand}
\end{figure}


\noindent
{\bf SA-based chiplet floorplanner.}
Our reach-aware floorplanner uses Sequence \textcolor{black}{Pairs}\textcolor{black}{s} \cite{MurataFNK96} 
to represent \textcolor{black}{a} spatial arrangement of chiplets in the netlist
and Simulated Annealing \cite{KirkpatrickGV83} to optimize the \textcolor{black}{objective} function \textcolor{black}{(Equation~\ref{eq:objective_sa})}. 
\textcolor{black}{Our annealer supports five solution perturbation (move) operators, 
each selected with equal probability (0.2)}.

\begin{itemize}[noitemsep,topsep=0pt,leftmargin=*]
    \item \textbf{Op1:} Swap two chiplets in the first sequence.
    \item \textbf{Op2:} Swap two chiplets in the second sequence.
    \item \textbf{Op3:} Swap two chiplets in both sequences.
    \item \textbf{Op4:} \textcolor{black}{Reshape a chiplet. Identify the chiplet with the highest
    reach-violation penalty and modify its shape using the resizing algorithm
    described in~\cite{ChenC06}. Figure~\ref{fig:resize-expand} (top) illustrates
    how a reach violation for the net connecting chiplets E and B (red dashed
    arrow) is resolved by aligning the right boundary of B with that of E.}
    \item \textbf{Op5:} \textcolor{black}{Expand (i.e., bloat) a chiplet. 
    Identify the chiplet with
    the highest reach-violation penalty and expand it into neighboring whitespace,
    where the expansion in each direction is determined by the extent required to
    eliminate the violation. Figure~\ref{fig:resize-expand} (bottom) shows how
    reach violations for nets connecting chiplets B--G and B--F are resolved by
    expanding B toward the top and right. This operator reduces reach-violation
    penalties at the cost of increasing} \textcolor{black}{the chiplet area and potentially the overall
    package area as well.}    
\end{itemize}

\noindent
\textcolor{black}{To enhance the performance of simulated annealing (SA), we adopt a {\em Go-With-the-Winners} (GWTW) strategy~\cite{Aldous94},
which enables \textcolor{black}{10} parallel SA {\em walkers} to independently
explore the solution space. Periodically, a synchronization phase is performed in which: (i) the best-performing threads are identified, (ii) their solutions are cloned to repopulate the thread pool and
(iii) all threads resume independent exploration. This strategy balances diversification and
intensification, improving convergence without incurring significant runtime overhead. We use \textcolor{black}{10}
threads in our experiments, a setting that is easily supported on standard server-class
machines.}\footnote{This number of threads is typical for modern multi-core servers and does not 
impose a significant hardware burden.}
\textcolor{black}{As discussed in Section \ref{sec:partition_refine}, \textcolor{black}{our} chiplet floorplanner operates in two modes: 
``standard'' and ``fast''. The standard mode performs 1,000,000 
perturbations with a cooling rate of 0.989, while the 
fast mode performs 10,000 perturbations with \textcolor{black}{an initial temperature of 0.0} 
(greedy). In our experiments, the ``standard mode'' is invoked after each FM or KL pass 
(default: 4 passes), totaling four invocations per run. 
The fast mode is triggered after each FM vertex move (default: 50\% of all vertices) 
and each KL vertex swap (default: 10\% of all vertices), resulting in \textcolor{black}{230 invocations per pass} 
in our largest testcase. These parameters \textcolor{black}{are} empirically selected to \textcolor{black}{maintain} a 
balance between runtime and solution quality.}

\section{Experimental Results}
\label{sec:results}

{\em ChipletPart} is implemented using approximately 
27K lines of C++ code, building 
upon implementations from \cite{BustanyGKKPW23}.
We use OpenMP~\cite{openmp} for parallelization, the Eigen 
library~\cite{eigen_lib} for spectral partitioning, 
codes from ~\cite{metis_repo} to run METIS, and Boost~\cite{boost}
for high precision arithmetic computations in the cost model.
Additionally, we have translated the Python-based cost model 
from ~\cite{Graening_CATCH,costmodelUCLA} into C++ and integrated 
it into our {\em ChipletPart} framework. The scripts and source 
code of {\em ChipletPart}, including its cost model implementation, are 
available in ~\cite{anonymousrepo}. We run all experiments on 
a Linux server with Intel 
Xeon E5-2690 CPU (48 threads) and 256GB RAM.

\subsection{Benchmarks and Baselines}
\label{subsec:benchmarks}

\textcolor{black}{
We evaluate four categories of testcases in this work: \textcolor{black}{{\em Waferscale}, {\em MemPool}, 
{\em Industrial Testcase}, and {\em Comparison Testcases}}. 
A common issue in \textcolor{black}{chiplet} partitioning research 
is that most testcases are too small to warrant practical chiplet partitioning; 
existing works often focus on very small designs, overemphasizing the impact of \textcolor{black}{I/O} and 
underemphasizing yield and packaging costs. To address this, we scale up the power 
and area of our baseline designs to sizes more representative of real commercial 
systems (e.g., Intel’s Ponte Vecchio~\cite{IntelPV}). We also scale the interconnects 
using Rent’s Rule,\footnote{Rent’s Rule~\cite{Landman71} states that the number of 
terminals $T$ in a logic block scales as $T = kC^p$, where $k$ is a constant, $C$ is the 
number of components and $p$ is the Rent parameter. Hence, if $C$ is scaled by a factor 
$s$, $T$ is scaled by $s^p$.} applying constants from~\cite{Singh1994} for microprocessor 
and memory blocks. This ensures that our scaled designs have a realistic number of \textcolor{black}{I/Os}, 
making them suitable for evaluating chiplet partitioning. 
}

\textcolor{black}{
Our first category of testcases (\emph{Waferscale}) is derived from 
the waferscale graph processor in \cite{Pal2021}. 
We extract \textcolor{black}{synthesized} IP block areas and a block-level netlist 
at 45\,nm, then apply area and power scaling to favor partitioning. 
The second category (\emph{MemPool}) is similarly extracted 
from \textcolor{black}{synthesized} blocks in \cite{Cavalcante2021} and likewise scaled from its \textcolor{black}{$45\text{nm}$} 
block-level netlist. 
The third category (\emph{Industry Testcase}) is based on a \textcolor{black}{$16\text{nm}$} 
design provided by \textcolor{black}{a semiconductor company}, which we 
also scale to a suitable size. 
Finally, the \emph{Comparison Testcases} \textcolor{black}{are} based on publicly available information for large commercial designs, 
\textcolor{black}{and} come from \cite{Li24}. 
\textcolor{black}{The original implementations
span multiple technology nodes, and these designs are sufficiently large that no
additional scaling is required.}}

\textcolor{black}{To ensure that routing congestion is not a concern in our example systems, 
we performed a simple analytical check based on estimated
routing capacity and required wires count. We considered both 
{\em escape routing} (breakout of signals from each chiplet on the substrate) 
and {\em global routing} (inter-chiplet connections across the substrate).
In all experiments, we considered a substrate with four routing layers and a 
routing pitch of \textcolor{black}{$1\mu$m}. This is similar to existing technologies \cite{Liao2014}.  
For escape routing, we verified whether the number of available routing tracks around 
each chiplet’s perimeter is sufficient to accommodate all connections to that die. 
For global routing, we examined the vertical slices across the substrate and 
confirmed that each slice contained enough routing tracks to support 
all inter-chiplet connections crossing that slice. 
Additionally, we used alternating routing directions across routing layers for global routing.
Under these considerations, we did not observe routing congestion in any of our testcases.}

Table \ref{tab:testcases} shows the specific testcase 
configurations that we use;
see also the design-specific discussion below. 
In the table, the \textit{WS} testcases are based on the 
{\em Waferscale} design, the \textit{MP} testcase is based on 
the {\em MemPool} design, \textcolor{black}{the \textit{TC} testcase is 
based on the {\em Industry} design, and \textit{EPYC} and \textit{GA100} 
are the {\em Comparison Testcases}.} 
The specific \textcolor{black}{design configurations} are explained as follows.

\begin{figure}[!htb]
    \centering
    \includegraphics[width=0.8\columnwidth]{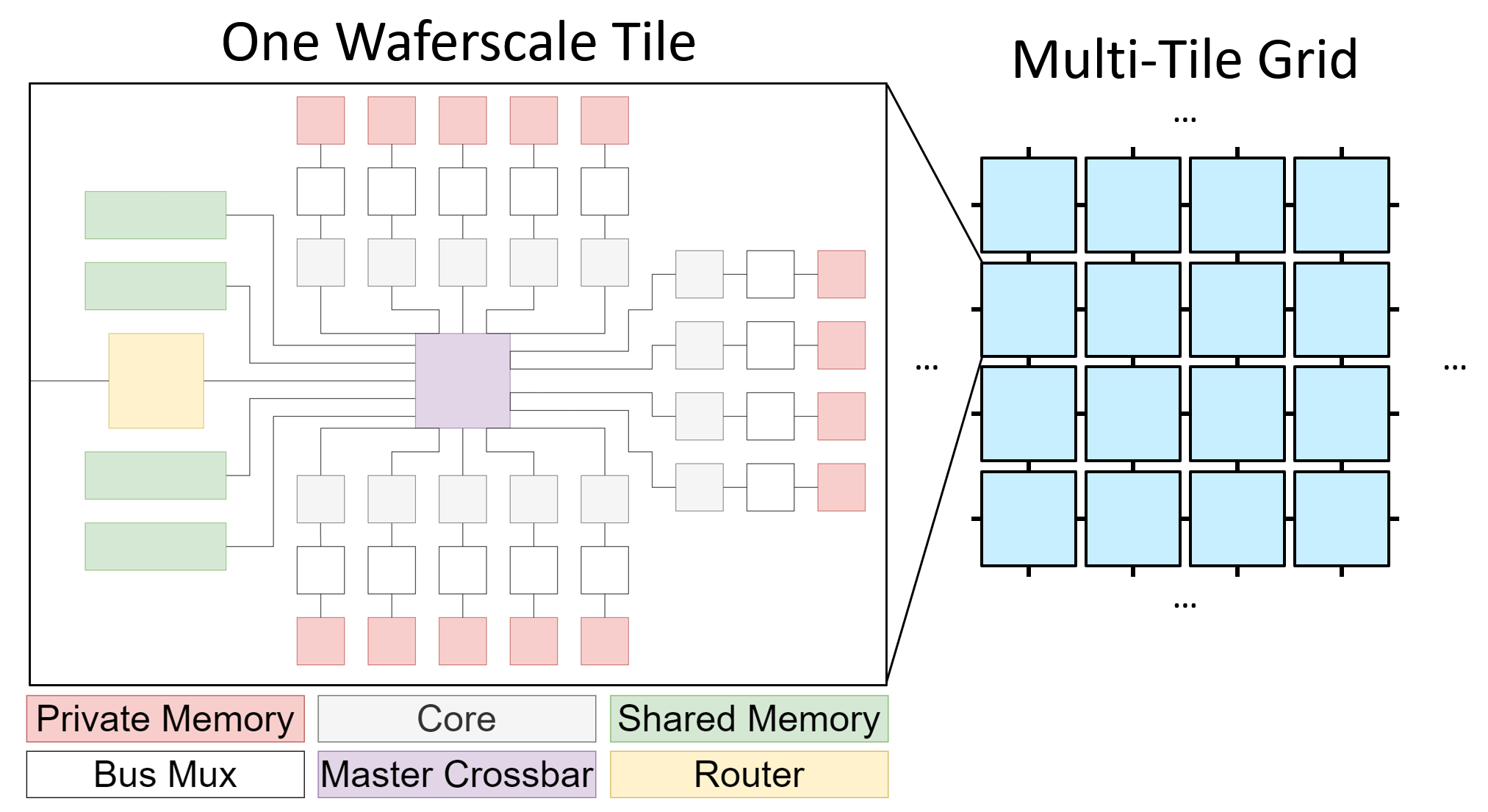}
    \Description{Image shows a block diagram of the waferscale testcase used in the paper. The image shows a tiled configuration where the blocks within each tile are arranged around a centralized crossbar.}
    \caption{\textcolor{black}{{\em Waferscale}} testcase. Left: example IP blocks in a tile. 
    Right: Connection of tiles in a grid.
    This testcase allows scaling of the number of tiles to match a 
    target design size for partitioning.}
    \label{fig:ws-testcase}
\end{figure}

\noindent
\textbf{{\em Waferscale Testcases.}} The \textcolor{black}{{\em Waferscale}} design is organized 
into a grid of tiles connected in a 2D 
mesh configuration with neighbor-to-neighbor communication 
between tiles. 
Each tile comprises 48 IP blocks. Those blocks include a router for 
tile-to-tile connections, a crossbar for intra-tile connections, \textcolor{black}{four} large 
shared memory blocks connected to the crossbar, and 14 cores each with 
associated bus and private memory blocks. 
Figure \ref{fig:ws-testcase} gives a system block diagram. 
We have implemented a
testcase generator (see \cite{anonymousrepo})
which allows us to choose the number of tiles in a given testcase, along 
with the number of cores and memories per tile.  Area and power scaling 
are also provided via the testcase generator. As noted, 
Table \ref{tab:testcases} shows the \textcolor{black}{{\em Waferscale}} configurations 
used in our case studies.

\begin{table}
\centering
\caption{\textcolor{black}{Benchmark characteristics. 
Note that MP and WS both use tile terminology; the others do not.}}
\label{tab:testcases}
\renewcommand{\arraystretch}{1.2} 
\setlength{\tabcolsep}{8pt} 
\scriptsize
\resizebox{1.00\columnwidth}{!} {
\begin{tabular}{@{}lccccc@{}}
\toprule
\textbf{Benchmark} & \textbf{\# Tiles} & \textbf{\# IP Blocks} & \textbf{Area ($mm^2$)} & \textbf{Area Scaling} & \textbf{Power Scaling} \\
\midrule
WS\textsubscript{1}  & 1    & 48  & 1582   ($45\text{nm}$) & 1600 & 1600 \\
WS\textsubscript{2}  & 2    & 96  & 3165   ($45\text{nm}$) & 1600 & 1600 \\
WS\textsubscript{3}  & 4    & 192 & 6330   ($45\text{nm}$) & 1600 & 1600 \\
WS\textsubscript{4}  & 8    & 384 & 12,660 ($45\text{nm}$) & 1600 & 1600 \\
MP                   & 16   & 40  & 494    ($45\text{nm}$) & 100  & 100  \\
TC\textsubscript{1} & N/A  & 14  & 567     ($16\text{nm}$) & 16   & 16   \\
TC\textsubscript{2} & N/A  & 17  & 567     ($16\text{nm}$) & 16   & 16   \\
EPYC                 & N/A  & 32  & 148    ($7\text{nm}$) & 1    & 1    \\
GA100                & N/A  & 180 & 548    ($7\text{nm}$)  & 1    & 1    \\
\bottomrule
\end{tabular}}
\end{table}


\noindent
\textbf{{\em MemPool Testcase.}} The full \textcolor{black}{{\em MemPool}} design 
consists of \textcolor{black}{four} \textcolor{black}{{\em MemPool}} {\em groups}, each containing 16 
\textcolor{black}{{\em MemPool}} {\em tiles}. These tiles each consist of cores and memory along 
with supporting logic. Note that while each tile contains \textcolor{black}{four} cores and 16 
memory banks, we do not split tiles for partitioning. For 
our \textcolor{black}{{\em MemPool}} testcase, we use a single \textcolor{black}{{\em MemPool}} group, i.e., taken at
just above the tile level of the hierarchy. Based on 16 tiles 
and 24 blocks related
to (remote, local, AXI etc.) interconnect, our block-level netlist 
for \textcolor{black}{{\em MemPool}} group
has 40 IP blocks to use in partitioning. A diagram is available 
in \cite{anonymousrepo}; 
see also Figure 3(a) in \cite{Cavalcante2021}.


\noindent
\textbf{\textcolor{black}{\emph{Industry Testcases.}}} 
\textcolor{black}{
We use a chip design from industry with parameters shown in 
Table \ref{tab:testcases}.\footnote{\textcolor{black}{The industry 
testcases are generated based on discussions with Analog Devices 
engineers \cite{ADI24}.}} Because the original design 
is relatively small, 
we scale the area and interconnect via Rent’s Rule to reach 
an overall design area 
of more than 400\,mm$^2$. This architecture follows a controller-target structure 
around a crossbar, where the crossbar constitutes 30\% of the total 
area of its connected blocks.
}



\begin{figure}
    \centering
    \includegraphics[width=0.9\linewidth]{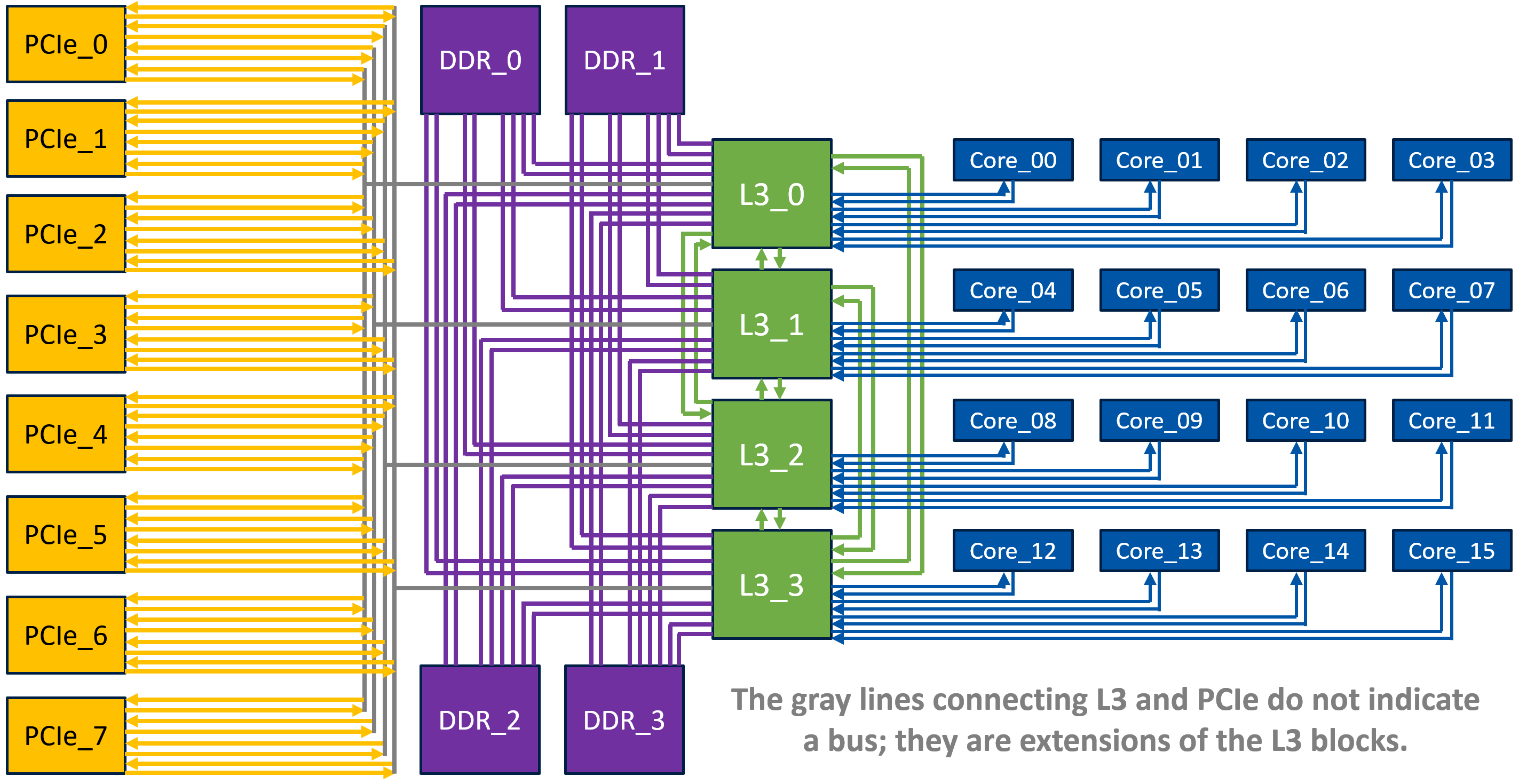}
    \Description{Image showing the blocks in the AMD EPYC 7282 testcase. Blocks are represented by rectangles and connections are shown with colored lines.}
    \caption{\textcolor{black}{AMD EPYC 7282 Testcase used in \cite{Li24}.}}
    \label{fig:epyc_7282}
\vspace{-0.1in}
\end{figure}

\noindent
\textbf{\textcolor{black}{{\em Comparison Testcases.}}}
\textcolor{black}{
We also evaluate on two commercial designs used 
by \textcolor{black}{{\em Chipletizer}}~\cite{Li24}. 
The first is based on AMD's \textcolor{black}{{\em EPYC}} 7282~\cite{epyc7282}, shown in 
Figure~\ref{fig:epyc_7282}. 
We additionally examine NVIDIA's \textcolor{black}{{\em GA100}} chip~\cite{ga100}, 
which powers the A100 GPU. 
The \textcolor{black}{{\em GA100}} design includes 128 simultaneous multiprocessor blocks, 
40 L2 blocks 
and multiple HBM controller blocks, totaling 180 blocks. 
We omit the \textcolor{black}{{\em GA100}} block diagram due to its size.
}

\noindent
\textbf{Metrics.} For simplicity, we assume iso-performance 
for our testcases, even across 
technology nodes; instead, we derive advantages in power from 
improvements in \textcolor{black}{I/O} count or changes in technology node. We also assume 
that connections between blocks in a chiplet will have the same speed as 
inter-chiplet connections, although the inter-chiplet connections will 
have higher power and area requirements due to large \textcolor{black}{I/O} drivers. 
\textcolor{black}{Design area is a major contributor to cost, and chip power contributes to chip cost via added bumps for current delivery that require additional area. For this reason, we use cost as a single metric for partitioning.}


\noindent
\textbf{Baselines.}
Since there is no open-source, cost-driven chiplet partitioner available, we compare against: 
(i) \emph{Min-cut partitioners} (\emph{hMETIS}~\cite{KarypisK96}, which is often used in chiplet flows~\cite{Kabir20}, and \emph{TritonPart}~\cite{BustanyGKKPW23}) applied to weighted hypergraphs, where each hyperedge is weighted by the corresponding I/O bandwidth;
(ii) the \emph{parChiplet} method from {\em Floorplet}~\cite{ChenLZZL23}; 
(iii) \emph{Chipletizer}~\cite{Li24}, for which we thank the authors for 
providing access to their code; and 
(iv) partitioning solutions generated by human engineers.


\begin{table}[htbp]
    \centering
    \caption{Partitioning costs and number of chiplets for hMETIS, TritonPart, Floorplet~\cite{ChenLZZL23}, Manual, Chipletizer~\cite{Li24}, $\text{ChipletPart}_{\text{No-FP}}$, and ChipletPart. Best cost per benchmark is in bold. We additionally report the geometric mean (GM) of absolute costs as well as the GM percentage improvement relative to the Manual baseline. Positive GM improvement indicates lower (better) cost on average. Chipletizer results are shown only for testcases where valid solutions were obtained. Because these cover a subset of the full design suite, GM values computed over six testcases are not directly comparable to the nine-testcase GM; hence, we denote Chipletizer’s GM cost (and improvement \%) as N/A.\protect\footnotemark}
    \label{tab:partitioning_costs_only}
    \renewcommand{\arraystretch}{1.2}
    \adjustbox{max width=\textwidth}{
        \begin{tabular}{|c|cc|cc|cc|cc|cc|cc|cc|}
            \hline
            \multirow{2}{*}{\textbf{Benchmark}} &
            \multicolumn{2}{c|}{\textbf{hMETIS}} &
            \multicolumn{2}{c|}{\textbf{TritonPart}} &
            \multicolumn{2}{c|}{\textbf{Floorplet}} &
            \multicolumn{2}{c|}{\textbf{Manual}} &
            \multicolumn{2}{c|}{\textbf{Chipletizer}} &
            \multicolumn{2}{c|}{\textbf{$\text{ChipletPart}_{\text{No-FP}}$}} &
            \multicolumn{2}{c|}{\textbf{ChipletPart}} \\
            \cline{2-15}
            & $|\mathcal{C}|$ & Cost  
            & $|\mathcal{C}|$ & Cost  
            & $|\mathcal{C}|$ & Cost  
            & $|\mathcal{C}|$ & Cost  
            & $|\mathcal{C}|$ & Cost  
            & $|\mathcal{C}|$ & Cost  
            & $|\mathcal{C}|$ & Cost  \\
            \hline
            WS\textsubscript{1} & 5 & 72.4   & 3 & 77.3   & 6 & 55.5   & 1 & 71.2   & 1 & 71.2   & 7 & 55.1   & 6 & \textbf{53.9} \\
            WS\textsubscript{2} & 8 & 138.1  & 7 & 143.2  & 4 & 150.3  & 2 & 145.5  & \text{Fail} & \text{Fail} & 6 & 122.3  & 8 & \textbf{123.4} \\
            WS\textsubscript{3} & 7 & 316.3  & 8 & 362.4  & 4 & 310.5  & 4 & 310.5  & \text{Fail} & \text{Fail} & 8 & 319.6  & 7 & \textbf{300.4} \\
            WS\textsubscript{4} & 8 & 1449.5 & 8 & 1564.2 & 5 & 1237.9 & 8 & 674.3  & \text{Fail} & \text{Fail} & 7 & 776.2  & 8 & \textbf{659.9} \\
            MP                  & 1 & 5.7    & 1 & 5.7    & 1 & 5.7    & 1 & 5.7    & 1 & 5.7    & 1 & 5.7    & 1 & \textbf{5.7}   \\
            TC\textsubscript{1} & 3 & 72.6   & 3 & 68.1   & 6 & 45.6   & 1 & 70.5   & 1 & 70.5   & 4 & 47.2   & 6 & \textbf{46.5} \\
            TC\textsubscript{2} & 3 & 73.9   & 3 & 71.1   & 7 & 46.5   & 4 & 49.4   & 1 & 74.9   & 6 & 48.7   & 7 & \textbf{46.2} \\
            EPYC                & 4 & 92.1   & 4 & 96.3   & 4 & 89.3   & 4 & 80.8   & 6 & 139.3  & 8 & 76.1   & 8 & \textbf{72.5} \\
            GA100               & 3 & 52.3   & 3 & 60.5   & 2 & 42.1   & 5 & 36.1   & 5 & 43.8   & 2 & 33.6   & 5 & \textbf{31.8} \\
            \hline
            \textbf{GM cost}
            & -- & \textbf{148.3}
            & -- & \textbf{171.0}
            & -- & \textbf{113.7}
            & -- & \textbf{114.9}
            & -- & \textbf{N/A}\footnotemark[17]
            & -- & \textbf{85.6}
            & -- & \textbf{82.3} \\
            \hline
            \textbf{GM impr. (\%)}
            & -- & \textbf{-21\%}
            & -- & \textbf{-26\%}
            & -- & \textbf{-2\%}
            & -- & \textbf{0\%}
            & -- & \textbf{N/A}\footnotemark[17]
            & -- & \textbf{+9\%}
            & -- & \textbf{\textbf{+13\%}} \\
            \hline
        \end{tabular}
    }
\end{table}

\footnotetext[17]{\textcolor{black}{For the six testcases where {\em Chipletizer} produces
solutions, the geometric mean cost (GM cost) is 48.5 vs.\ 33.9 for {\em ChipletPart}, with {\em ChipletPart} achieving
a geo-mean cost improvement (GM imprv.) of 30.1\%.}}

\begin{table}[htbp]
\centering
\caption{\textcolor{black}{Chiplet count and I/O-feasibility for hMETIS, TritonPart, Floorplet~\cite{ChenLZZL23}, Manual, Chipletizer~\cite{Li24}, $\text{ChipletPart}_{\text{No-FP}}$, and ChipletPart. ``Feas'' indicates I/O-feasibility (\cmark: feasible, \xmark: not feasible; ``Fail'' indicates that no valid solution was obtained).
We also report the number of successful runs for each category.}}
\label{tab:partitioning_feas_only}
\renewcommand{\arraystretch}{1.2}
\adjustbox{max width=\textwidth}{
\begin{tabular}{|c|cc|cc|cc|cc|cc|cc|cc|}
\hline
\multirow{2}{*}{\textbf{Benchmark}} &
\multicolumn{2}{c|}{\textbf{hMETIS}} &
\multicolumn{2}{c|}{\textbf{TritonPart}} &
\multicolumn{2}{c|}{\textbf{Floorplet}} &
\multicolumn{2}{c|}{\textbf{Manual}} &
\multicolumn{2}{c|}{\textbf{Chipletizer}} &
\multicolumn{2}{c|}{\textbf{$\text{ChipletPart}_{\text{No-FP}}$}} &
\multicolumn{2}{c|}{\textbf{ChipletPart}} \\
\cline{2-15}
& $|\mathcal{C}|$ & Feas  
& $|\mathcal{C}|$ & Feas  
& $|\mathcal{C}|$ & Feas  
& $|\mathcal{C}|$ & Feas  
& $|\mathcal{C}|$ & Feas  
& $|\mathcal{C}|$ & Feas  
& $|\mathcal{C}|$ & Feas  \\
\hline
WS\textsubscript{1} & 5 & \xmark & 3 & \cmark & 6 & \xmark & 1 & \cmark & 1 & \cmark & 7 & \xmark & 6 & \cmark \\
WS\textsubscript{2} & 8 & \xmark & 7 & \xmark & 4 & \cmark & 2 & \cmark & \text{Fail} & \text{Fail} & 6 & \xmark & 8 & \cmark \\
WS\textsubscript{3} & 7 & \xmark & 8 & \xmark & 4 & \cmark & 4 & \cmark & \text{Fail} & \text{Fail} & 8 & \xmark & 7 & \cmark \\
WS\textsubscript{4} & 8 & \xmark & 8 & \xmark & 5 & \xmark & 8 & \cmark & \text{Fail} & \text{Fail} & 7 & \xmark & 8 & \cmark \\
MP                  & 1 & \cmark & 1 & \cmark & 1 & \cmark & 1 & \cmark & 1 & \cmark & 1 & \cmark & 1 & \cmark \\
TC\textsubscript{1} & 3 & \cmark & 3 & \cmark & 6 & \xmark & 1 & \cmark & 1 & \cmark & 4 & \cmark & 6 & \cmark \\
TC\textsubscript{2} & 3 & \cmark & 3 & \cmark & 7 & \xmark & 4 & \cmark & 1 & \cmark & 6 & \xmark & 7 & \cmark \\
EPYC                & 4 & \cmark & 4 & \cmark & 4 & \cmark & 4 & \cmark & 6 & \xmark & 8 & \xmark & 8 & \cmark \\
GA100               & 3 & \cmark & 3 & \cmark & 2 & \cmark & 5 & \cmark & 5 & \xmark & 2 & \cmark & 5 & \cmark \\
\hline
\textbf{\#Successes (out of 9)}
& \multicolumn{2}{c|}{4}
& \multicolumn{2}{c|}{6}
& \multicolumn{2}{c|}{5}
& \multicolumn{2}{c|}{9}
& \multicolumn{2}{c|}{4}
& \multicolumn{2}{c|}{3}
& \multicolumn{2}{c|}{\textbf{9}} \\
\hline

\end{tabular}}
\end{table}

\subsection{Validations of Chiplet Partitioning}
\label{subsec:validations_of_par}

Given that previous works use traditional min-cut partitioners in 2.5D flows, we directly compare {\em ChipletPart} with leading partitioners {\em hMETIS} and {\em TritonPart}. We also compare to the {\em Floorplet} floorplan-aware chiplet partitioner \cite{ChenLZZL23}, {\em Chipletizer}~\cite{Li24}, and {\em manual} partitions. For the \textcolor{black}{{\em Waferscale}} testcases based on \cite{Pal2021} we compare the memory-compute partition in that work with a per-tile partitioning as the manual partition. Overall, our validations span (i) validation of chiplet cost-driven partitioning, (ii) validation of floorplan-awareness, (iii) validation of heterogeneous technology-awareness, and (iv) analyses of hyperparameter sensitivities and runtime. 

\subsubsection{Validation of chiplet cost-driven partitioning}
\label{subsubsec:valid_cost_driven_part}
We evaluate {\em ChipletPart} against the baselines listed in the previous section: {\em hMETIS}, {\em TritonPart}, {\em Floorplet}, {\em Chipletizer}, and manually derived partitions. Our validation is thus divided into three parts, detailed below.


\noindent
\textbf{Comparison with netlist partitioners.}
We evaluate the chiplet cost-driven partitioning capabilities of {\em ChipletPart} by benchmarking against leading netlist partitioners {\em hMETIS} and {\em TritonPart}, using the latter's default parameter settings.\footnote{For {\em hMETIS}, the default settings are Nruns = 10, CType = 1, RType = 1, Vcycle = 1, Reconst = 0, and seed = 0~\cite{KarypisAKS99}. For {\em TritonPart}, the default parameter settings are $thr\_coarsen\_hyperedge\_size\_skip = 200$, $coarsening\_ratio = 1.6$, $max\_moves = 60$, and $num\_coarsen\_solutions = 3$~\cite{BustanyGKKPW23}.} We use the testcases listed in Table~\ref{tab:testcases} for our evaluations, with results presented in Tables~\ref{tab:partitioning_costs_only} and~\ref{tab:partitioning_feas_only}. Table~\ref{tab:partitioning_costs_only} presents chiplet cost comparisons and Table~\ref{tab:partitioning_feas_only} presents \textcolor{black}{floorplan-feasibility} comparisons. For both {\em hMETIS} and {\em TritonPart}:
\begin{itemize}[noitemsep,topsep=0pt,leftmargin=*]
    \item We set an imbalance factor of 5\% and define our input set of partitions as ${K} = {1, 2, 3, ..., 10}$.
    \item For each $K$, we execute ten runs of the partitioner, yielding a total of 100 partitioning solutions.\footnote{Our studies show that longer partitioner runs do not improve the quality of results.}
    \item We evaluate each partitioning solution using our cost model (Secs.~\ref{sec:cost_driver} and \ref{sec:power_model}) and report the solution with lowest chiplet cost.
    \item Since {\em hMETIS} and {\em TritonPart} are not technology-aware, for a fair comparison, we enforce homogeneity by running {\em ChipletPart} with only the \textcolor{black}{7 $\text{nm}$} technology node. Consequently, the cost model evaluates all partitioning solutions using the \textcolor{black}{7 $\text{nm}$} technology node. The floorplanner is run in the standard mode (Section~\ref{sec:floorplan}).
\end{itemize}

\noindent
Results are presented in Tables~\ref{tab:partitioning_costs_only} and~\ref{tab:partitioning_feas_only}. Columns $|C|$, $Cost$, and $Feas$, respectively indicate the number of chiplets, the evaluated cost of the chiplet solution, and whether the solution satisfies the reach constraints. {\em ChipletPart} consistently produces solutions with better cost than {\em hMETIS} and {\em TritonPart} with improvements of up to 55\% and 58\% respectively. The geometric mean improvements over {\em hMETIS} and {\em TritonPart} are 16\% and 20\%, respectively. This result shows the advantages of {\em ChipletPart}'s cost-driven partitioning over traditional min-cut partitioners. Furthermore, {\em ChipletPart} consistently produces floorplan-feasible solutions, whereas standard netlist partitioners often fail to meet the reach constraints.\footnote{{\em hMETIS} and {\em TritonPart} optimize cutsize and return partitions with better cutsize compared to {\em ChipletPart} (albeit with worse {chiplet} cost).}


\noindent
\textbf{Comparison with previous work.}
We compare {\em ChipletPart} with the {\em parChiplet} partitioner
from {\em Floorplet}~\cite{ChenLZZL23}, and {\em Chipletizer}~\cite{Li24}. 
Here, we set the {\em Floorplet} parameters as $|\mathcal{C}|_{\min} = 1$, $|\mathcal{C}|_{\max} = 10$ and $rr = 1.0$. 
All partitioning solutions are evaluated using our cost model with the \textcolor{black}{7 $\text{nm}$} technology node and the floorplanner is run in standard mode.
Tables~\ref{tab:partitioning_costs_only} and~\ref{tab:partitioning_feas_only} 
present these results. Compared to {\em Floorplet}, {\em ChipletPart} achieves up to a 47\% cost improvement, with a geometric mean improvement of $6\%$. While {\em Floorplet} produces a solution with slightly lower cost on the \textcolor{black}{$TC_1$} testcase, the solution violates reach constraints. In contrast, {\em ChipletPart} consistently generates floorplan-feasible solutions with better cost, highlighting its strength in delivering high-quality and physically realizable chiplet partitions. For {\em Chipletizer}, we use the default settings from the original implementation, with {\em production volume}~\cite{Li24} set to 500{,}000 units. {\em Chipletizer} fails to produce a 
solution on three of nine testcases \textcolor{black}{($WS_1, WS_2, WS_3, WS_4$)} due to scalability limitations. Across the remaining testcases, {\em ChipletPart} achieves up to 48\% lower cost\footnote{A more effective way to run {\em Chipletizer} would be to integrate it with our cost
model. Since this is beyond the scope of our current work, we leave it for future research.} compared to {\em Chipletizer}, while consistently generating I/O-feasible solutions. For example, on {\em GA100}, both frameworks produce a five-chiplet solution, but {\em ChipletPart} achieves a 27\% cost reduction and satisfies I/O-feasibility, whereas the {\em Chipletizer} solution violates feasibility.


\noindent
\textbf{Comparison with human baselines.}
We generate human baselines for each testcase as follows. 
For the \emph{Waferscale} ($WS_1, WS_2, WS_3, WS_4$) designs, which are tile-based, the human partitioning solution aligns directly with these tiles. For the \emph{red}{4}-way partition with the same blocks as $TC_1$ but with a 4-way split crossbar. For the \emph{EPYC} testcase (Figure~\ref{fig:epyc_7282}), we cluster each set of cores with its corresponding L3 --- distributing other blocks to maintain regularity and balanced area --- resulting in four partitions each containing  \textcolor{black}{two} PCIe blocks,  \textcolor{black}{one} DDR,  \textcolor{black}{one} L3, and  \textcolor{black}{four} cores. Similarly, for \emph{GA100}, we create five chiplets: four partitions each holding 32 simultaneous multiprocessors and 10 L2 blocks, and one partition for the remaining blocks.\footnote{Our human baselines do not necessarily reflect partitions chosen by industry. Instead, we split the blocks to generate relatively even partitions with small cutset and high regularity as an example of what the human engineer might decide to do.}

Similar to the previous sections, all partitioning solutions are evaluated using our cost model with the $7\text{nm}$ technology node and the floorplanner is run in standard mode. Tables~\ref{tab:partitioning_costs_only} and~\ref{tab:partitioning_feas_only} present the results. {\em ChipletPart} generates solutions that are up to $34\%$ better than the manual baselines with a geometric mean improvement of $13\%$. 
\textcolor{black}{While the human baselines are all floorplan-feasible, they are not cost-optimal.
These results indicate that a black-box partitioner such as \emph{ChipletPart}
can generate higher-quality solutions than manual partitioning strategies.}

\subsubsection{Validation of floorplan-awareness}
\textcolor{black}{We evaluate the impact of floorplan-awareness in \emph{ChipletPart}.
Tables~\ref{tab:partitioning_costs_only} and~\ref{tab:partitioning_feas_only}
compare solutions generated with and without floorplan-awareness enabled,
denoted as \emph{ChipletPart} and \emph{ChipletPart$_\text{No-FP}$}
respectively. Disabling floorplan-awareness can occasionally yield lower
nominal chiplet cost (e.g., $WS_2$), but it frequently produces solutions that
violate interconnect reach constraints. In particular, six out of nine testcases
result in infeasible solutions when floorplan-awareness is disabled. Beyond
feasibility, enabling floorplan-awareness improves chiplet cost by up to 15\%,
with a geometric mean improvement of 4\%.} 
This is because our cost model incorporates floorplan-level information when evaluating partition quality --- thus, floorplan-aware solutions better align with the true cost objective. These observations highlight the importance of incorporating floorplan-awareness: while it may slightly increase nominal cost in some cases, it helps to ensure that all generated solutions are physically realizable.

\begin{figure}
    \centering
    \includegraphics[width=0.9\linewidth]{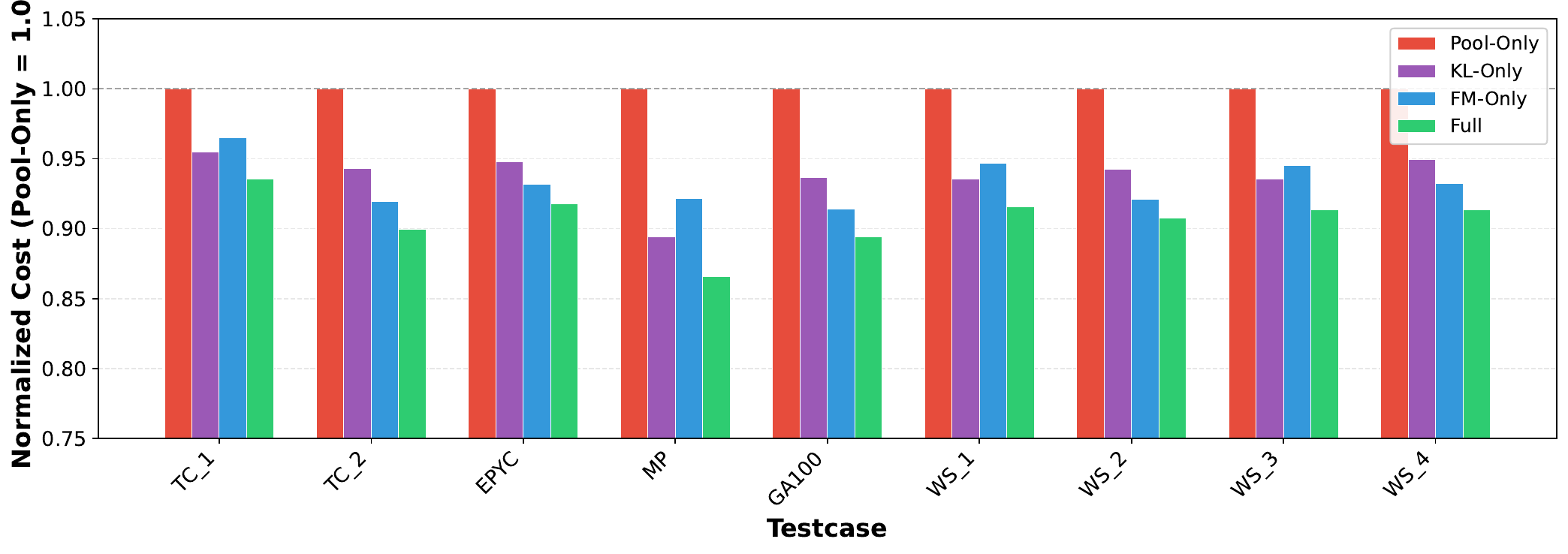}
    \Description{A bar chart showing the performance of running ChipletPart with some of the features turned off.}
    \caption{Impact of initial partitioning and refinement 
    in {\em Core-ChipletPart}. {\em Pool-Only} denotes {\em Core-ChipletPart}
    with no refinement. {\em KL-Only} denotes {\em Core-ChipletPart} with
    KL refinement only. {\em FM-Only} denotes {\em Core-ChipletPart} with
    FM refinement only.}
    \label{fig:ablation_core}
\vspace{-0.1in}
\end{figure}

\subsubsection{Impact of initial partitioning and refinement}
\label{subsubsec:impact_partitioning}
To quantify the contribution of each component in {\em Core-ChipletPart}, 
we conduct 
an ablation study comparing four configurations: 
(i) the {\em full} framework i.e., {\em Core-ChipletPart}; 
(ii) initial partitioning {\em pool-only} with no refinement;
(iii) initial partitioning pool with FM refinement only; and (iv) initial partitioning pool with KL refinement only. 
Figure~\ref{fig:ablation_core} presents the results across all 
testcases from Table~\ref{tab:testcases}.
The full {\em Core-ChipletPart} achieves an average cost reduction of 9.3\% compared 
to the \textcolor{black}{{\em pool-only}} baseline, with improvements ranging from 6.4\% to 13.4\% 
across different designs. When examining the individual refinement strategies, 
\textcolor{black}{{\em FM-only}} achieves 6.4\% average improvement while \textcolor{black}{{\em KL-only}} achieves 5.7\% average 
improvement over the pool-only baseline. 
Notably, neither refinement strategy
consistently dominates the other: \textcolor{black}{{\em FM-only}} 
outperforms \textcolor{black}{{\em KL-only}} on five out of ten 
testcases, while \textcolor{black}{{\em KL-only}} performs better on the remaining four testcases. 
This complementary behavior demonstrates that the two refinement algorithms optimize
different aspects of the partitioning solution. FM excels at move-based local
optimization while KL provides effective swap-based refinement for certain design
topologies.
Importantly, the {\em full} {\em Core-ChipletPart} consistently outperforms both 
individual refinement strategies across all testcases, achieving an additional 
2-3\% improvement over the better of \textcolor{black}{{\em FM-only}} or \textcolor{black}{{\em KL-only}} in each case. 
We believe that these results validate our integrated approach: the diverse initial
partition pool provides high-quality starting points, while the 
application of FM and KL refinement generates further improvement
that neither algorithm achieves independently.

\subsubsection{Validation of heterogeneous technology-awareness}
\label{subsubsec:hetero}
We evaluate benefits of incorporating heterogeneous technology awareness on all nine testcases (Table~\ref{tab:testcases}) and a technology list comprising \textcolor{black}{7 $\text{nm}$, 
10 $\text{nm}$, and 14 $\text{nm}$} nodes. 
\textcolor{black}{Table~\ref{tab:hetero_awares}
reports chiplet cost reductions achieved by enabling technology awareness,
relative to the best solutions obtained under homogeneous technology
integration for each node.
Specifically, we first run \emph{ChipletPart} using
a single technology node per testcase (homogeneous integration) to obtain
baseline partitioning solutions for each technology. We then run
\emph{ChipletPart} with technology awareness enabled (i.e., allowing the use
of all three technology nodes) and compare the resulting solutions.}
Our results show that
heterogeneous technology assignment can reduce system cost by up to
43\%, with geometric mean cost reductions of 7\%, 15\%, and 15\% when compared against homogeneous solutions at \textcolor{black}{7 $\text{nm}$},
\textcolor{black}{10 $\text{nm}$}, and \textcolor{black}{14 $\text{nm}$}, respectively.\footnote{\textcolor{black}{To 
further validate the effectiveness of our GA, we implemented an
enumeration-based framework that exhaustively explores all
$\sum_{k=1}^{K_{\text{max}}} \binom{k + m - 1}{m - 1}$ possible assignments. Our experiments show that the GA achieves up to $7\times$ speedup with less than 1\% degradation in chiplet cost compared to the enumeration-based framework.}}
Note, that all solutions generated by {\em ChipletPart} are \textcolor{black}{floorplan-feasible}.

\begin{table}
\centering
\caption{Impact of heterogeneity. ``Homogeneous''  denotes homogeneous integration cost. ``Heterogeneous'' denotes heterogeneous integration cost, chiplet tech distribution, and {\em ChipletPart} runtime. A single invocation of {\em Core-ChipletPart's
runtime is $\sim$$5\%$} of {\em ChipletPart}'s runtime. Runtimes reported are the CPU time.}
\label{tab:hetero_awares}
\renewcommand{\arraystretch}{1.2}
\setlength{\tabcolsep}{2.8pt}
\begin{tabular}{|l|ccc|c|ccc|c|}
\hline
\multirow{2}{*}{\textbf{Bench.}} &
\multicolumn{3}{c|}{\textbf{Homogeneous}} &
\multicolumn{5}{c|}{\textbf{Heterogeneous}} \\
\cline{2-9}
 & 7nm & 10nm & 14nm & Cost & \#7nm & \#10nm & \#14nm & Runtime (s) \\
\hline
WS\textsubscript{1}  & 53.9  & 48.9  & 42.1  & \textbf{\textcolor{black}{41.7}} & \textcolor{black}{2} & 0 & 4 & 401 \\
WS\textsubscript{2}  & 123.4 & 112.7 & 109.4 & \textbf{87.8} & 1 & 0 & 7 & 1421 \\
WS\textsubscript{3}  & 300.4 & 310.4 & 286.3 & \textbf{254.4} & 3 & 0 & 5 & 3240 \\
WS\textsubscript{4}  & 659.9 & 662.4 & 676.1 & \textbf{656.2} & 7 & 1 & 0 & 5624 \\
MP                  & 5.7   & 6.9   & 8.8   & \textbf{5.7}  & 1 & 0 & 0 & 180 \\
TC\textsubscript{1}  & 46.5  & 52.4  & 56.4  & \textbf{39.8} & 3 & 0 & 2 & 160 \\
TC\textsubscript{2}  & 46.2  & 54.2  & 64.1  & \textbf{39.3} & 4 & 0 & 2 & 89 \\
EPYC                & 72.5  & 86.2  & 94.6  & \textbf{65.1} & 3 & 0 & 2 & 349 \\
GA100               & 31.8  & 40.5  & 55.2  & \textbf{31.6} & 3 & 1 & 0 & 493 \\
\hline
\end{tabular}
\end{table}

\begin{figure}
    \centering
    \includegraphics[width=\columnwidth]{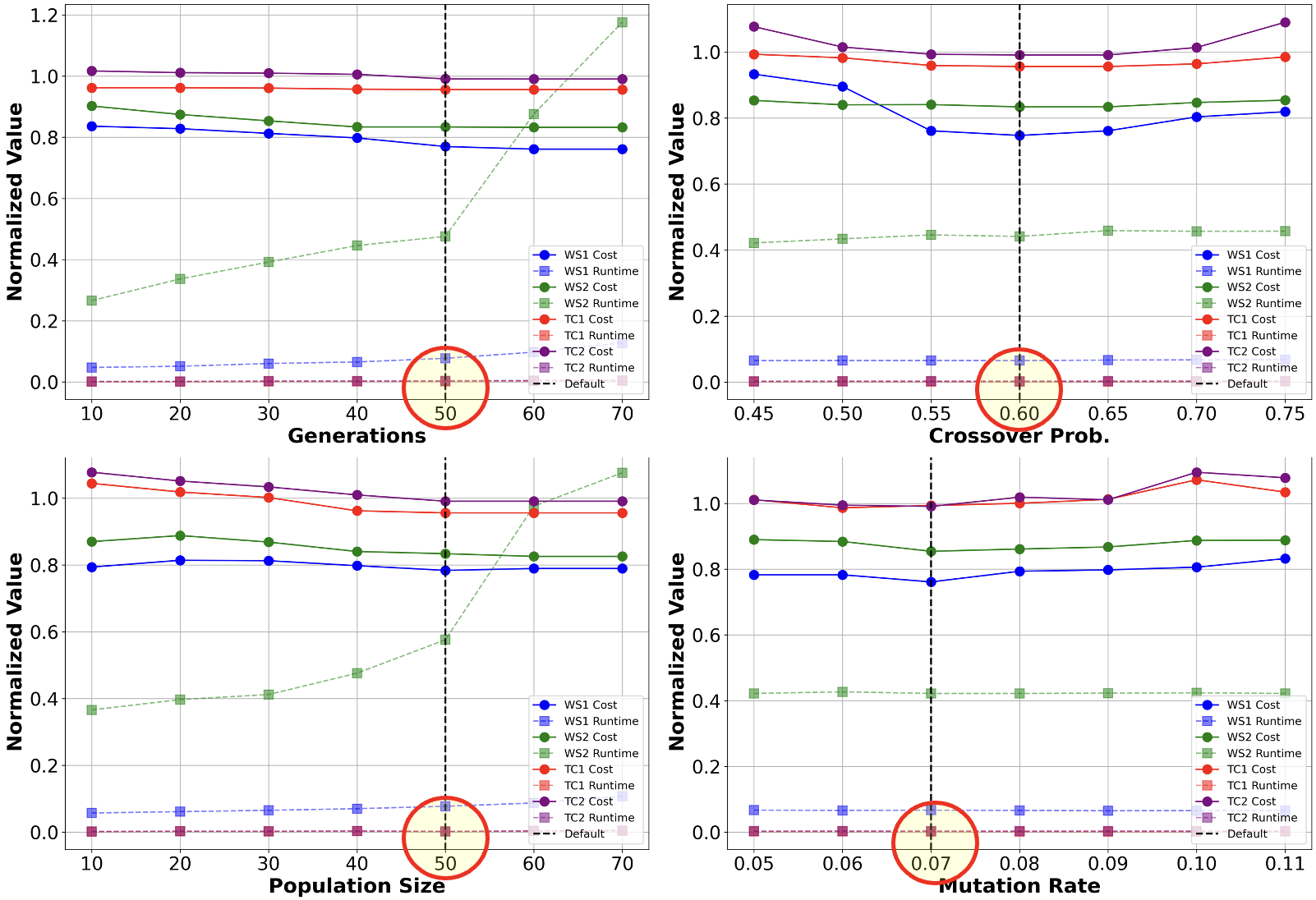}
    \Description{Image includes 4 plots of the impact of hyperparameters on results.}
    \caption{Hyperparameters sweep across benchmarks. \textcolor{black}{S}olid lines denote normalized cost and dashed lines denote normalized runtime. The default hyperparameter choice is marked with a vertical dashed line and circled.}
    \label{fig:hyperparam_sweep}
\vspace{-0.1in}
\end{figure}

\begin{figure}
    \centering
    \includegraphics[width=0.6\linewidth]{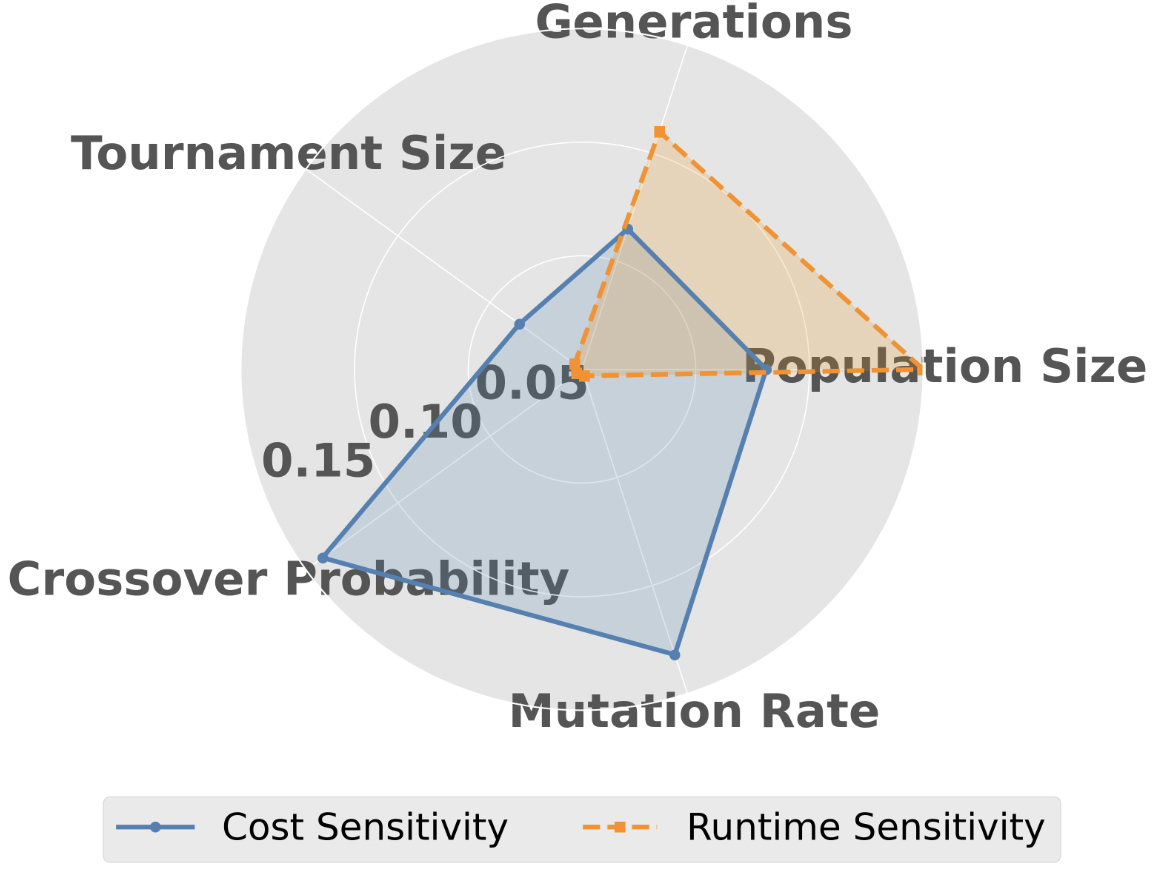}
    \Description{Image of a spider chart showing the sensitivity to various hyperparameters.}
    \caption{Sensitivity analysis of hyperparameters. Larger \textcolor{black}{distances from the plot center} indicate greater influence.}
    \label{fig:hyperparam_sensitivity}
\end{figure}

\subsubsection{\textcolor{black}{Hyperparameter exploration and sensitivity analysis}}
\label{subsubsec:hyperparam_explore}
{\em ChipletPart}'s \textcolor{black}{genetic algorithm}  relies on four key hyperparameters: 
(i) total population size $tot_{pop}$; 
(ii) maximum number of generations $\Psi$; 
(iii) mutation probability $p_m$; and 
(iv) crossover probability $p_c$ (cf. Algorithm~\ref{alg:overall_chipletpart}).\footnote{{\em ChipletPart} never returns a chiplet solution with more than 8 chiplets, so we set $K_{max}$ to 8.}
We validate the default values of these hyperparameters via an empirical study using testcases 
$WS_1$, $WS_2$, $TC_1$, and $TC_2$ from Table~\ref{tab:testcases}. 
We use the following technology nodes for this
study: \textcolor{black}{7 $\text{nm}$}, \textcolor{black}{10 $\text{nm}$,} and \textcolor{black}{14~$\text{nm}$}, and run {\em ChipletPart} on
the four aforementioned testcases. 
Figure~\ref{fig:hyperparam_sweep} 
presents the normalized chiplet costs (normalized to the cost of 
{\em manual} solutions) and the normalized runtime (represented as ``Normalized Value'' in Figure~\ref{fig:hyperparam_sweep}), normalized to a constant value of $10000$ seconds.\footnote{$10000$ seconds is chosen for normalization to allow for comparison with about three hours of runtime.}
In our experiments, we systematically vary each hyperparameter while keeping the others fixed at their default values. 
For each configuration, we measure the resulting chiplet cost and runtime across the testcases. 
Our observations from Figure~\ref{fig:hyperparam_sweep} indicates that the default hyperparameter settings achieve a 
balanced trade-off between partitioning quality and runtime efficiency.

Figure~\ref{fig:hyperparam_sensitivity} summarizes the
sensitivity of {\em ChipletPart} to its
hyperparameters. We measure sensitivity as the normalized variance in cost and
runtime across the sweep range for each hyperparameter. Notably, crossover and mutation probability have the highest influence on cost, while
 population size and number of generations dominate runtime variation.
 Tournament size has relatively low sensitivity for both metrics. These
 results indicate that careful tuning of crossover and mutation parameters is
 essential for chiplet solution quality, while population size control primarily
 affects runtime overhead.

\begin{table}
\centering
\caption{Runtime breakdown of {\em ChipletPart}.}
\label{tab:runtime_profile}
\renewcommand{\arraystretch}{1.2}
\setlength{\tabcolsep}{8pt}
\begin{tabular}{|l|c|}
\hline
\textbf{Component} & \textbf{Runtime share (\%)} \\
\hline
SA-based floorplanner & 67 \\
Partition refinement (chiplet cost model)  & 28 \\
Partition pool         & 2  \\
Partition pruning      & 1 \\
\textcolor{black}{File} I/O \textcolor{black}{handling}                    & 1  \\
\hline
\end{tabular}
\end{table}

\subsection{Runtime remarks}
\label{subsec:runtime_profile}
{\em ChipletPart} is implemented entirely in C++, including a 
high-quality C++ translation of the Python-based cost model from~\cite{costmodelUCLA}. 
Our optimized C++ translation achieves approximately a ${5}\times$ speedup compared to 
the original Python version. 
Empirical results indicate that {\em ChipletPart}, when integrated with the translated 
C++ cost model, attains over a $100\times$ speedup relative to an integration 
based on the Python cost model. 
Additionally, our \textcolor{black}{open-sourced} implementation is fully parallelized using \textcolor{black}{{\em OpenMP}~\cite{openmp}}. 
These improvements make {\em ChipletPart} 
readily adaptable for any chiplet-based design flow. 
\textcolor{black}{
We present a detailed runtime breakdown of {\em ChipletPart} in Table~\ref{tab:runtime_profile}. 
Approximately 67\% of the execution time is spent \textcolor{black}{on} floorplanning, 
while 28\% is spent on chiplet cost evaluation. 
Further details on {\em ChipletPart}'s runtime are provided in Table~\ref{tab:hetero_awares}.
} 

\noindent
\textcolor{black}{\textbf{Scalability.} For large designs such as
\textcolor{black}{$WS_4$}, our runtime remains under \textcolor{black}{two} hours.
To improve scalability
further, hierarchical decomposition techniques that exploit structural regularity in
the netlist, along with aggressive solution pruning and potential 
usage of surrogate models for faster evaluation\textcolor{black}{,} can be employed.}

\subsection{Exploring Bayesian Optimization}
\label{subsec:bayesian_opt}

\textcolor{black}{Bayesian optimization (BO) is a powerful 
framework for optimizing expensive, black-box functions,
particularly when gradients are unavailable and each 
function evaluation incurs significant computational
overhead~\cite{shahriari2015taking}. Unlike exhaustive or
random search strategies, BO constructs a probabilistic
\textit{surrogate model} of the objective function and uses 
an \textit{acquisition function}~\cite{gan2021acquisition} 
to balance exploration and exploitation. These properties 
make BO especially suitable for problems with complex constraints, cost landscape, and costly evaluations.
To provide another quality-runtime tradeoff point, we have also explored the \textcolor{black}{potential} 
use of BO for chiplet partitioning. We jointly optimize (i) the assignment of IP blocks to chiplets, and (ii) the selection of technology nodes
per chiplet to minimize total system cost. 
Directly optimizing the partitioning function \(\phi : \mathcal{S} \rightarrow \mathcal{C}\) is infeasible due to 
the combinatorial explosion in the search space: for \(k\)
chiplets and \(|\mathcal{S}|\) blocks, there 
are \(k^{|\mathcal{S}|}\) possible assignments. 
Such a high-dimensional, discrete space is ill-suited to BO,
which relies on continuous, low-dimensional representations~\cite{papenmeier2025}.
To address this, we use spectral embeddings to derive a continuous relaxation of the partitioning problem~\cite{chung1997spectral}. We construct a normalized Laplacian matrix~\cite{merris1994laplacian} 
from the netlist connectivity graph and compute its $d$ nontrivial eigenvectors. These eigenvectors 
define a mapping:
\[
    \Phi_{\text{embed}}: \mathcal{S} \rightarrow \mathbb{R}^d, \quad s \mapsto \mathbf{e}_s
\]
where $s \in \mathcal{S}$ is an IP block, and $\mathbf{e}_s \in
\mathbb{R}^d$ is its coordinate in the $d$-dimensional spectral
embedding space. Blocks that are tightly coupled (i.e., frequently communicate or share many nets) are 
embedded closer together. This transformation preserves the structural information of the netlist in a form 
more amenable to clustering~\cite{bustany2022specpart}. Rather than directly optimizing $\phi$, we allow BO to propose a set of $k$ centroids $\{\mathbf{c}_1, \ldots, \mathbf{c}_k\}$ in $\mathbb{R}^d$. Each block is then assigned to its nearest centroid:
\begin{equation}
    \phi(s) = \arg\min_{j \in \{1, \dots, k\}} \| \mathbf{e}_s - \mathbf{c}_j \|_2
\end{equation}
This transforms the partitioning problem from a high-dimensional discrete search into a continuous 
optimization over $k \cdot d$ real-valued variables. Moreover, this centroid-based assignment 
strategy naturally produces geometrically coherent partitions and enables integration with 
downstream refinement algorithms such as FM or KL.}

\textcolor{black}{In addition to partitioning, the BO
framework must also determine the technology assignment
$\omega(c) \in \mathcal{T}$ for each chiplet $c \in
\mathcal{C}$. Technology assignments are categorical variables,
which are challenging to surrogates due to their non-ordinal nature. 
To mitigate this, we
encode each technology assignment using a one-hot vector:
\[
    \mathbf{t}_i \in \{0, 1\}^m, \quad \text{where } m = |\mathcal{T}|, \quad \sum_{j=1}^{m} \mathbf{t}_i[j] = 1
\]
Here, $\mathbf{t}_i$ represents the technology assigned to chiplet $c_i$. 
This encoding allows the BO surrogate model to treat technology assignment as a 
structured subspace within a continuous optimization landscape. Each candidate solution evaluated by BO is represented as a vector $x$:}
\begin{itemize}
    \item \textcolor{black}{The number of partitions $k$ (integer-valued).}
    \item \textcolor{black}{$k$ centroids in $d$-dimensional spectral space: $\mathbf{c}_1, \ldots, \mathbf{c}_k$.}
    \item \textcolor{black}{One-hot technology encodings $\mathbf{t}_1, \ldots, \mathbf{t}_k$.}
\end{itemize}

\begin{table}
\centering
\caption{\textcolor{black}{Comparison between \textit{ChipletPart-GA} and \textit{ChipletPart-BO}.
{\em ChipletPart-GA} denotes {\em ChipletPart} using GA as the optimizer. {\em ChipletPart-BO}
denotes {\em ChipletPart} using BO as the optimizer.
Runtimes reported are the CPU time.}}
\label{tab:ga_bo_comparison}
\renewcommand{\arraystretch}{1.2}
\setlength{\tabcolsep}{3pt}
\begin{tabular}{|l|ccccc|ccccc|}
\hline
\multirow{2}{*}{\textbf{Bench.}} & \multicolumn{5}{c|}{\textbf{ChipletPart-GA}} & \multicolumn{5}{c|}{\textbf{ChipletPart-BO}} \\
\cline{2-11}
& Cost & \#7nm & \#10nm & \#14nm & Runtime (s)
& Cost & \#7nm & \#10nm & \#14nm & Runtime (s) \\
\hline
WS\textsubscript{1}  & 41.7 & 2 & 0 & 4 & 401 & 41.7 & 1 & 0 & 4 & 848 \\
WS\textsubscript{2}  & \textbf{87.8} & 1 & 0 & 7 & 1421 & 97.1 & 0 & 3 & 5 & 4262 \\
WS\textsubscript{3}  & 254.4 & 3 & 0 & 5 & 3240 & \textbf{251.4} & 4 & 0 & 4 & 9839 \\
WS\textsubscript{4}  & 656.2 & 7 & 1 & 0 & 5624 & \textbf{640.2} & 6 & 1 & 1 & 12149 \\
MP                  & 5.7 & 1 & 0 & 0 & 180 & 5.7 & 1 & 0 & 0 &  350 \\
TC\textsubscript{1}  & \textbf{39.8} & 3 & 0 & 2 & 160 & 42.1 & 4 & 1 & 1 & 1137 \\
TC\textsubscript{2}  & \textbf{39.3} & 4 & 0 & 2 & 89 & 40.2 & 5 & 1 & 2 & 959 \\
EPYC                & \textbf{65.1} & 3 & 0 & 2 & 349 & 67.8 & 4 & 4 & 0 & 1263 \\
GA100               & 31.6 & 3 & 1 & 0 & 493 & 31.6 & 3 & 1 & 0  & 1685 \\
\hline
\end{tabular}
\end{table}

\textcolor{black}{Formally, $x = [k, \mathbf{c}_1, \ldots, \mathbf{c}_k, \mathbf{t}_1, \ldots, \mathbf{t}_k]$ and the dimensionality of $x$ is: $\dim(x) = 1 + k \cdot d + k \cdot m$. 
The BO engine constructs a surrogate model $\hat{\Phi}(x)$ to approximate the 
true cost $\Phi(x)$ and uses an acquisition function $\alpha(x)$ to guide exploration.
We use a Gaussian Process (GP)~\cite{frazier2018tutorial} surrogate for small datasets ($< 50$ IP blocks) and Random Forest~\cite{zhang2021prediction} for larger datasets. Our 
acquisition function comprises Expected Improvement~\cite{gan2021acquisition} and Upper Confidence Bound~\cite{gan2021acquisition} metrics. Once BO picks a {\em candidate} $x$, it undergoes the following evaluation pipeline:}

\begin{enumerate}
    \item \textcolor{black}{\textit{Partition decoding}: Assign each block $s \in \mathcal{S}$ to its nearest
    centroid to obtain partition $\phi$.
    \item \textit{FM refinement}: Improve the partition using {\em ChipletPart}'s FM
    refinement. }
    \item \textcolor{black}{\textit{Technology decoding}: Map each one-hot vector $\mathbf{t}_i$ to its 
    corresponding technology node $\omega(c_i)$.
    \item \textit{Cost evaluation}: Call the {\em ChipletPart}'s cost model to evaluate the cost $\Phi(\phi, \omega)$, and floorplan checks.}
    \item \textcolor{black}{\textit{Feasibility check}: If the solution is infeasible, i.e., does not satisfy the reach
    constraints, 
    return a large penalty value $M \gg 1$: $\Phi(\phi, \omega) = M$. This ensures that the 
    BO explores the landscape where the solutions are always floorplan-feasible.}
\end{enumerate} 

\textcolor{black}{The result is fed back to update the surrogate model, and the
process is repeated until a termination criterion is met (we
use a fixed limit of 200 iterations). We compare our 
implemented BO framework with our {\em ChipletPart}'s GA framework.
Table~\ref{tab:ga_bo_comparison} presents a cost and runtime comparison of the two 
optimizers --- \textcolor{black}{{\em ChipletPart-GA} and {\em ChipletPart-BO}}. The results show that {\em ChipletPart-BO} 
can produce lower-cost solutions on the larger designs\textcolor{black}{, $WS_3$ and $WS_4$}, 
where it achieves cost reductions of up to 5.3\% and 2.4\%, respectively. However, the benefits come 
at a significant computational cost. On average, {\em ChipletPart-BO} incurs a runtime overhead of 
2$\times$–4$\times$ compared to {\em ChipletPart-GA}, due to its expensive surrogate model updates. 
In small-to-medium benchmarks such as \textcolor{black}{$MP$, $TC_2$, and $GA100$}, both optimizers 
achieve comparable costs, but BO remains considerably slower.} \textcolor{black}{Thus, GA and BO provide a 
quality-runtime trade-off.}

\noindent
\textbf{BO vs. GA trade-offs.} Our results show that {\em ChipletPart-GA} provides the best overall
quality/runtime tradeoff for chiplet partitioning: it converges quickly,
is robust across benchmarks, and integrates naturally into a broader CAD
flow. In contrast, {\em ChipletPart-BO} incurs significantly higher
runtime due to surrogate-model construction and acquisition-function
optimization, but can yield modestly better solutions on benchmarks whose
spectral embeddings exhibit a smoother cost landscape
(e.g., \textcolor{black}{$WS_3$, $WS_4$}). We recommend {\em ChipletPart-GA} 
for routine use or time-constrained flows, while {\em ChipletPart-BO} 
can potentially serve as a
compute-intensive alternative that may improve quality on select, 
more challenging testcases.

    \section{Additional Case Studies}
    \label{sec:case_studies}
    
    We investigate how different technology parameters 
    affect {\em ChipletPart}'s solutions. 
    To accommodate an arbitrarily complex cost model, {\em ChipletPart} employs a 
    black-box partitioning method rather than a mathematical programming approach. 
    This design choice enables a broader exploration of the parameter space 
    compared to other partitioners. 
    To illustrate these benefits, we present case studies that examine effects of 
    altering \textcolor{black}{the chiplet} cost model parameters. 
    
    
    \begin{figure}[htbp]
        \centering
        \begin{subfigure}{1.0\columnwidth}
            \includegraphics[width=\linewidth]{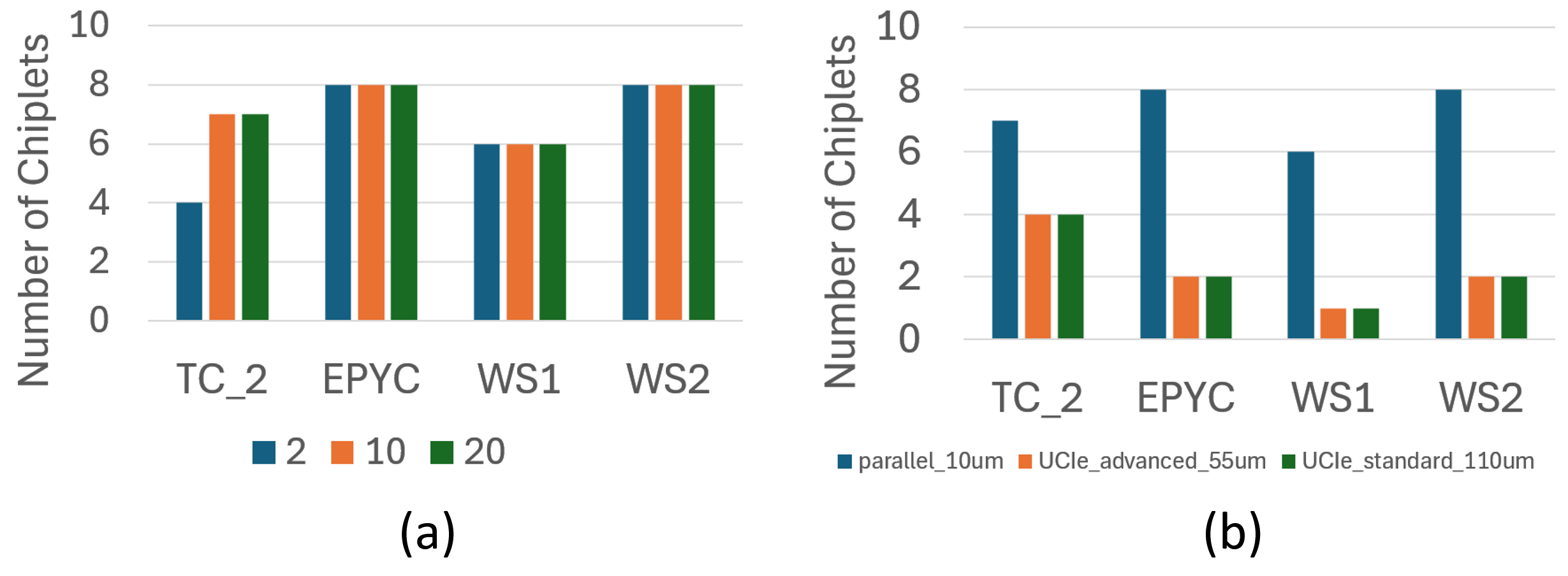}
        \end{subfigure}
        \Description{Image containing two bar charts showing different testcases with varied reach constraints and I/O types.}
        \caption{
        (a) Reach limit vs. \#chiplets. The partitioning solutions
        are generated with {\em ChipletPart} using reach values of 
        \textcolor{black}{2 $\text{mm}$, 10 $\text{mm}$, and 20 $\text{mm}$}. (b) \textcolor{black}{I/O} type vs. \#chiplets.}
        \label{fig:case_studies}
    \vspace{-0.1in}
    \end{figure}



    
    \noindent
    \textbf{\textcolor{black}{I/O} type \textcolor{black}{and reach}.} \textcolor{black}{
    Chiplet systems often use reduced-size \textcolor{black}{I/O} cells that trade driver 
    strength and \textcolor{black}{electrostatic discharge} (ESD) 
    protection for a smaller form factor. However, limited \textcolor{black}{I/O} reach makes it challenging to 
    generate feasible solutions. A \textcolor{black}{2 $\text{mm}$} reach may 
    only allow direct connections between adjacent chiplets, while a 
    larger reach (\textcolor{black}{e.g.,} \textcolor{black}{20 $\text{mm}$}) increases flexibility but also adds to system cost 
    by increasing chiplet area. To explore this, we: (i) analyze 
    the impact of \textcolor{black}{I/O} reach on chiplet partitioning for a \textcolor{black}{{\em parallel}} \textcolor{black}{I/O} testcase; and (ii) compare \textcolor{black}{the {\em parallel} I/O testcase} with UCIe \textcolor{black}{{\em advanced}} and UCIe 
    \textcolor{black}{{\em standard}} definitions \cite{UCIe1p0}, which 
    have different cell sizes and reach. 
    The parallel \textcolor{black}{I/O} has \textcolor{black}{2 $\text{mm}$} reach, while the UCIe cases have \textcolor{black}{10 $\text{mm}$} reach}.
    
    \textcolor{black}{Figure~\ref{fig:case_studies}(a) shows that increasing reach enables a 
    greater number of chiplets in the partitioning solution, while a short \textcolor{black}{2 $\text{mm}$} 
    reach forces more blocks to remain together in fewer chiplets. 
    As shown in Figure~\ref{fig:reach_example}, longer I/O reach values enable 
    connections that are impossible under shorter reach limits. A reach of 
    \textcolor{black}{2 $\text{mm}$} effectively restricts connectivity to nearest neighbors, which forces more blocks 
    to remain colocated and increases overall cost. 
    This effect is clearly illustrated in 
    the \textcolor{black}{$TC_2$} testcase: with a \textcolor{black}{2 $\text{mm}$} reach,
    the resulting solution has a 
    cost of 46.3, whereas increasing the reach to \textcolor{black}{10 $\text{mm}$} produces a lower-cost solution 
    (cost 43.4), corresponding to a 6.3\% reduction.
    Figure~\ref{fig:case_studies}(b) compares the same \textcolor{black}{{\em parallel}} \textcolor{black}{I/O} testcase with UCIe 
    interfaces, showing that UCIe’s large \textcolor{black}{I/O} cells favor fewer chiplets, while the parallel 
    interface with smaller \textcolor{black}{I/O}s and bump pitches enables finer partitioning.} 
    \textcolor{black}{For example, this effect is evident in the \textcolor{black}{$EPYC$} testcase. We verified the 
    impact of \textcolor{black}{I/O} type and bump pitch by evaluating the cost of a fixed chiplet 
    partitioning solution, the {\em parallel\_10um} solution, under the other \textcolor{black}{I/O} configurations. 
    When evaluated with the {\em parallel\_10um} \textcolor{black}{I/O} configuration, the solution has a cost of 
    73.1, whereas evaluating the same partitioning solution under the 
    {\em UCIe\_advanced\_55um} and {\em UCIe\_standard\_110um} configurations increases the 
    cost to 74.1 and 74.5, respectively. Because this additional cost arises solely 
    from the 
    larger \textcolor{black}{I/O} cell sizes and pitches, the solutions optimized for 
    {\em UCIe\_advanced\_55um} and {\em UCIe\_standard\_110um} reduce the total inter-chiplet 
    connection bandwidth by merging blocks into fewer chiplets, thereby avoiding the higher \textcolor{black}{I/O} 
    cost.}
    
    
    \noindent
    \textbf{Mixed cost/power objective function.} {\em ChipletPart} can 
    simultaneously optimize cost and power using a weighted 
    sum of chiplet cost (Equation \ref{eq:part_cost}) and  
    power as its objective function. 
    Table \ref{tab:chiplets_coefficients} 
    shows results for various cost and power coefficient combinations. 
    For this experiment, we use the \textcolor{black}{$WS_1$ and $TC_1$} 
    testcases, with all chiplets implemented 
    in \textcolor{black}{7 $\text{nm}$} technology. \textcolor{black}{As the power coefficient increases, the solutions 
    favor fewer chiplets since large I/O drivers for inter-chiplet connections 
    incur higher power consumption.}
    
    \begin{table}[htbp]
    \centering
    \caption{\textcolor{black}{Effects of different cost and 
    power coefficient combinations. 
    Cost$_N$ and Power$_N$ refer to normalized cost and power, respectively.}}
    \label{tab:chiplets_coefficients}
    \begin{tabular}{cc|ccc|ccc}
    \toprule
    \multicolumn{2}{c|}{\textbf{Coefficients}} & \multicolumn{3}{c|}{\textbf{$WS_1$}} & \multicolumn{3}{c}{\textbf{$TC_1$}} \\
    \cmidrule(lr){1-2} \cmidrule(lr){3-5} \cmidrule(lr){6-8}
    \textbf{Cost} & \textbf{Power} & \textbf{$|C|$} & \textbf{Cost$_N$} & \textbf{Power$_N$} & \textbf{$|C|$} & \textbf{Cost$_N$} & \textbf{Power$_N$} \\
    \midrule
    1    & 0   & \textcolor{black}{6} & \textbf{1.00} & \textbf{1.00} & \textcolor{black}{6} & \textbf{1.00} & \textbf{1.00} \\
    0.75 & 0.25 & \textcolor{black}{6} & \textcolor{black}{1.02} & \textcolor{black}{0.97} & 4 & \textcolor{black}{1.03} & 0.97 \\
    0.5  & 0.5  & 4 & \textcolor{black}{1.06}  & \textcolor{black}{0.95} & 2 & \textcolor{black}{1.04}  & \textcolor{black}{0.95} \\
    0.25 & 0.75 & 4 &  \textcolor{black}{1.07} & \textcolor{black}{0.93} & 2 &  \textcolor{black}{1.07} & \textcolor{black}{0.93} \\
    0    & 1    & 2 &  \textcolor{black}{1.10}  & \textcolor{black}{0.92} & 1 &  \textcolor{black}{1.08}  & \textcolor{black}{0.91} \\
    \bottomrule
    \end{tabular}
    \vspace{-0.1in}
    \end{table}

\section{Conclusion}
\label{sec:conclusion}

We have proposed {\em ChipletPart}, a novel chiplet 
partitioning framework designed to  tackle the challenges of partitioning 
large, 2.5D heterogeneously integrated systems into chiplets. 
{\em ChipletPart} leverages an advanced cost model, is power-aware, and uses 
a genetic algorithm to simultaneously assign technologies to chiplets, 
all while consistently returning I/O-feasible solutions that honor the driving 
reach of inter-chiplet \textcolor{black}{I/O} drivers. 
\textcolor{black}{Experimental results demonstrate that {\em ChipletPart} 
outperforms classical netlist partitioners, 
{\em Floorplet}~\cite{ChenLZZL23}, and human baselines.
Additional studies illuminate dependencies of 
partitioning outcomes on the underlying packaging and 
\textcolor{black}{I/O} technology.}
We have also explored Bayesian optimization as an 
alternative search strategy. \textcolor{black}{Although Bayesian optimization can sometimes generate higher-quality solutions than the genetic algorithm, it incurs significantly higher runtime
overhead.} 


\textcolor{black}{While 
{\em ChipletPart} is designed as a cost-driven partitioning framework, its modular, black-box 
optimization approach promotes future integration of more complex objective functions that consider 
power, performance and thermal factors. Extending the current objective function to include additional factors remains a direction for future work.  \textcolor{black}{Considering detailed \textcolor{black}{I/O} planning} is \textcolor{black}{another} direction
for future work. 
Detailed \textcolor{black}{I/O} planning that supports pass-through connections, multiple \textcolor{black}{I/O} types on a die, \textcolor{black}{``gas station''} buffers~\cite{CoskunEJKMNS20}, buffer chiplets, or other similar architectural considerations will bring significantly increased problem complexity. \textcolor{black}{Evaluation with routing-congested designs is another topic of interest.} 
We also aim to enhance {\em ChipletPart} by identifying 
repeated structures in the netlist, a capability that 
could significantly improve scalability for larger \textcolor{black}{designs}. 
Beyond 2.5D, adapting {\em ChipletPart} \textcolor{black}{to 3D-stacked} technologies, where vertical interconnects, TSV constraints, and
tier locality introduce new partitioning objectives, is an
opportunity for future work. Incorporating detailed thermal and power
delivery models can potentially enable co-optimization for reliability,
temperature, and system-level performance.}
Finally, we are exploring the use of alternate optimizers, such as 
\textcolor{black}{Integer Linear Programming} (ILP) and \textcolor{black}{Reinforcement Learning} (RL). We also plan to integrate 
{\em ChipletPart} with OpenROAD~\cite{OpenROAD} to promote 
engagement in benchmarking and further development.

\begin{acks}
    \textcolor{black}{This work was supported in part by Samsung AI Center, Intel Corporation, DARPA/SRC CHIMES JUMP 2.0 center, NSF, and the CDEN center.}
\end{acks}

\bibliographystyle{IEEEtran}
\bibliography{references,IEEEabrv}

\end{document}